\documentclass[intlimits,twoside,a4paper]{article}
\usepackage[cp1251]{inputenc}

\usepackage[eqsecnum]{cmpj3}

\issue{2024}{27}{2}{23601}
\doinumber{10.5488/CMP.27.23601}

\title[Hamiltonian limited valence model]%
{Hamiltonian limited valence model for liquid polyamorphism}
\author[S. V. Buldyrev]
{S. V. Buldyrev\orcid{0000-0002-3008-1070}\refaddr{label1, label2}\thanks{E-mail: \email{buldyrev@verizon.net}}}
\addresses{
	\addr{label1} Department of Physics, Yeshiva University, New York, NY 10033, USA
	\addr{label2} Department of Physics, Boston University, MA 02215, USA  
}


\date{Received January 26, 2024, in final form March 05, 2024}

\begin{document}

\maketitle

\begin{abstract}
Liquid--liquid phase transitions have been found experimentally or by computer simulations in many compounds such as water, hydrogen, sulfur, phosphorus, carbon, silica, and silicon. Limited valence model implemented via event-driven molecular dynamics
algorithm provides a simple generic mechanism for the liquid--liquid phase transitions in all these diverse cases. Here, we introduce a variant of the limited valence model with a well defined Hamiltonian, i.e., a unique algorithm by which the potential energy of the system of particles
can be computed solely from the coordinates of the particles and is thus equivalent to a complex multi-body potential. We present several examples of the model which can be used to reproduce liquid--liquid phase
transition in systems with maximum valence $z=1$ (hydrogen), $z=2$ (sulfur)
and $z=4$ (water), where $z$ is the maximum number of bonds an atom is allowed to have. For $z=1$, we find a set of parameters for which the system has a liquid--liquid and an isostructural solid--solid critical points. For $z=4$, we find a set of parameters for which the phase diagram resembles that of water with a wide region of negative thermal expansion coefficient (density anomaly) extending into the metastable region of negative pressures. The limited valence model can be modified to forbid not only too large valences but also too low valences. In the case of sulfur, we forbid the formation of monomers, thus restricting the valence $v$ of an atom to be within an interval $1=v_{\rm min}\leqslant v\leqslant v_{\rm max}\equiv z=2$.
\printkeywords 
%
\end{abstract}

\section{Introduction}
Typically, pure substances may be found with only one gaseous or liquid state, while their solid state may exist in various polymorphic crystalline forms. The existence of two distinct liquid forms in a single component substance is more unusual since liquids lack the long-range order common to crystals. Yet, the existence of multiple amorphous liquid states in a single component substance, a phenomenon known as ``liquid polyamorphism'' \cite{Debenedetti_Water_1998,Stanley_Liquid_2013,Anisimov_Polyamorphism_2018,Tanaka_Liquid_2020}, has been observed or predicted in a wide variety of substances, such as superfluid helium \cite{Vollhardt_He_1990,Schmitt_He_2015}, high-pressure hydrogen \cite{Ohta_H_2015,Zaghoo_H_2016,McWilliams_H_2016,Norman_Review_2021,Fried_Hydrogen_2022}, sulfur \cite{Henry_Sulfur_2020}, phosphorous \cite{Katayama_Phos_2000,Katayama_Phos_2004}, carbon \cite{Glosli_Liquid_1999}, silicon \cite{Sastry_Silicon_2003,Beye_Silicon_2010,Vasisht_Solicon_2011,Sciorino_Silicon_2011}, silica \cite{Saika_Silica_2000,Lascaris_Silica_2014}, selenium and tellurium \cite{Tsuchiya_SeTe_1982,Brazhkin_SeTe_1999}, and cerium \cite{Cadien_Ce_2013}. Liquid polyamorphism is also highly plausible in deeply supercooled liquid water \cite{Angell_TwoState_1971,Angell_Amorphous_2004,Stanley_Liquid_2013,Anisimov_Polyamorphism_2018,Tanaka_Liquid_2020,Poole_Water_1992,Debenedetti_Water_1998,Holten_Water_2012,Holten_Water_2014,Gallo_Water_2016,Biddle_Water_2017,Caupin_Thermodynamics_2019,Duska_Water_2020}. 

The limited valence model introduced by Cummings and Stell \cite{Cummings1984} in 1984 has been used extensively in condensed matter physics \cite{Stell1993,Speedy1994,Speedy1996,Zaccarelli2005,Moreno2005,Zaccarelli2006,Kalyuzhnyi_2007,Kalyuzhnyi2007,Smallenburg2013,Rescic2016,Jamnik2017,Kalyuzhnyi2017,Stepanenko2019,Kalyuzhnyi2023}. 
More recently, it has been shown that a variant of limited valence model can reproduce liquid--liquid phase transitions (LLPT) in some cases listed above \cite{Shumovsky2022,Shumovsky2023}. The main idea of the model is that
the system consists of identical atoms interacting via spherically symmetric potentials. 
Depending on the number of atoms within a certain distance $w_b$ away from a given atom, this atom is assigned a unique type or valence, $v$:
$v=0$ with no neighbors within this distance, $v=1$ with one neighbor, $v=2$ with two neighbors and so on. If an atom has
more than $z$ neighbors, where $z$ is the limited valence, the potential energy of the system is assigned to be infinity and hence such a configuration cannot exist. After the types of atoms are determined by counting their neighbors, the potential energy is computed using a table of pair-wise potentials $U_{ij}(r)$ defined as the potential energy of the pair of atoms of types $i$ and $j$ separated by distance $r$. Similarly, we can forbid not only too large valences but also too low valences assigning atoms with $v<v_{\rm min}$ an infinite potential energy. The model can be straightforwardly implemented by the Monte-Carlo method.

This approach does not contradict the quantum understanding of inter-atomic interactions which suggests that for a fixed position of ions, the solution
of the Schrodinger equation for electrons has a ground state with the energy which depends in a very complex way on the position of the ions and, thus, does not allow simple parametrization of the force field in terms of the ion coordinates. Recently these difficulties were addressed by the machine learning \cite{Behler2021} in which the force field is learned from an extensive data base of sample solutions for typical ionic configuration. However, machine learning does not provide us with a simple physical interpretation of the inter-atomic interactions. The maximum valence model, on the other hand, provides a simple  cartoon interpretation of
this mechanism if we postulate that the changes in the density in a  very small vicinity of atoms limited to a length of a covalent bond or a hydrogen bond leads to drastic effects on longer range inter-atomic interactions.  In the case of hydrogen, the pioneering quantum calculations of Wigner and Huntington \cite{Wigner1935} suggested that at high density the energy of the metallic lattice is lower than the energy
of the molecular lattice. In other words, at high densities, H$_2$, dimers disassociate due to a steric effect; i.e., the electron shells of H$_2$ dimers are getting larger than the intermolecular
distances. This steric repulsion is expected to happen not only for the case of hydrogen but also for other molecular liquids with higher valences. Similar arguments may be applied to the case of water. It was
found experimentally that, at high densities, when the fifth neighbor is being pushed into the first coordination shell of a molecule, hydrogen bonds bifurcate and the energy of the
bifurcated bonds is roughly half of the energy of the straight bonds~\cite{Giguere1984}. Essentially, this implies that the shells of the atoms with coordination number 4 become impenetrable for a fifth intruder at low temperatures.

Historically, the limited valence model was implemented using
a reaction scheme in the discrete molecular dynamics (DMD) algorithm~\cite{Sitges}. DMD which is also known as event driven molecular dynamics was first applied in 1959 to a system of hard spheres~\cite{Alder1959}.  The simplest case is the model with $z=1$~\cite{Shumovsky2023},
which mimics the dimerization of metallic hydrogen into molecular hydrogen when the pressure is lowered. The model also reproduces a LLPT in sulfur for $z=2$~\cite{Shumovsky2022}, which is associated with the polymerization at high pressure~\cite{Henry_Sulfur_2020}, and a LLPT for $z=4$~\cite{Shumovsky2023},
which mimics the hypothesized LLPT in the four coordinated liquids such as water and silica and also reproduces the region of density anomaly known to exist in these liquids. 

In the previous publications~\cite{Shumovsky2022,Shumovsky2023} using the DMD algorithm, the Hamiltonian requirement for the limited valence model was omitted. This omission, obviously contradicts the regular definition of statistical mechanics, since in this case the potential energy of the system depends not
only on the configurational space but also on the trajectory of the system in
this space, thus making the definition of the entropy ambiguous. This problem also
arises in a directional variant of the limited valence model based on the
Kern-Frenkel model of patchy colloids~\cite{Kalyuzhnyi_2007} in which a pair of particles with overlapping patches may or may not have a bond, but if the patches are so wide that a patch overlaps with multiple patches of other particles, only one bond per patch is allowed. This is solved by introducing random reassignment of bonds among pairs of particles for which their geometrical
position allows them to have a bond, so that only at most one bond per patch can exist at each moment of time. The reassignment is produced with a probability proportional to the Boltzmann factor of the bond formation, which ensures correct evaluation of entropy via thermodynamic integration.

In this paper, we make a Hamiltonian formulation of the model by simply forbidding all other particles, except the bonded ones, to be closer to each other than 
the bond length, $w_z$.  We reexamine the results obtained previously~\cite{Shumovsky2022,Shumovsky2023} for $z=1$, $z=2$, and $z=4$ and see how the Hamiltonian condition alters the results.

The paper is organized as follows. In section \ref{sec:DMD} we define the Hamiltonian limited valence model and describe its in detail. In section \ref{sec:z=1} we compare the results of the Hamiltonian variant of the model with the previously studied non-Hamiltonian variant~\cite{Shumovsky2023} for $z=1$ and study how the LLPT changes with the width of the repulsive shoulder $w_z$ in the dimerization transition ($z=1$). In section \ref{sec:z=4} we study the same question
for $z=4$ and show that for the wide shoulder $w_z\approx 1.34$, the phase diagram resembles that of water. In section \ref{sec:z=2}
we study the problem of polymerization for $z=2$ as function of the bond
potential strength and find the values at which the positions of the LLPT and liquid--gas phase transition (LGPT) resemble those in sulfur.

\section{The model}
\label{sec:DMD}

\begin{figure}[t]
    \centering
    \includegraphics[width=0.90\linewidth]{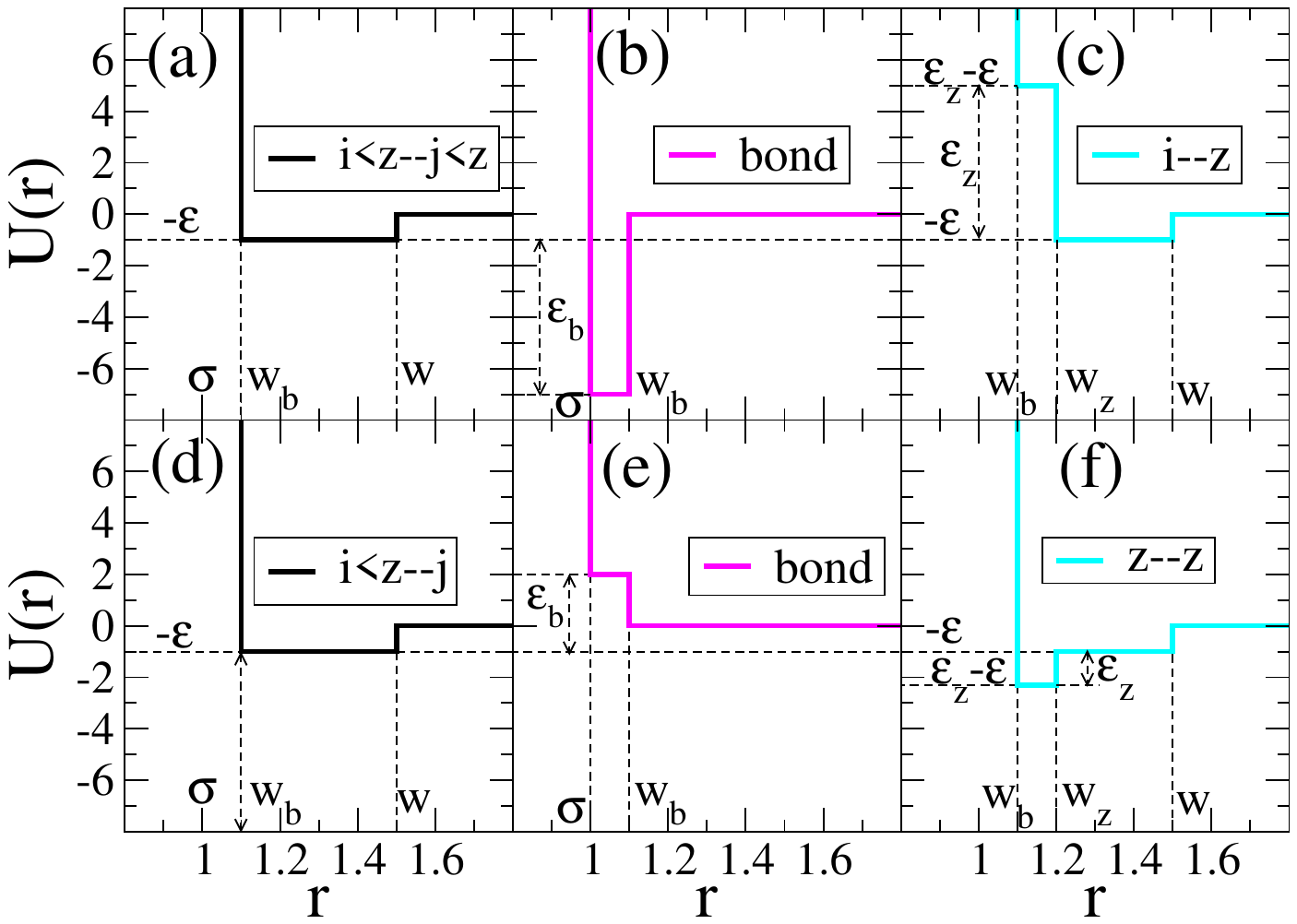}
    \caption{(Colour online) Panels (a, b, c). Pairwise potentials $U_{ij}$ which we use to model hydrogen ($z=1$) and water ($z=4$). Atoms interact with the square well potential of width $w>w_b>\sigma$ and energy $-\epsilon$ [black line, panel (a)] if their types are $i<z$ and $j<z$. If they come closer than the bond width $w_b$, they change their types to $i+1$ an $j+1$ and they form a bond  [magenta line, panel (b)] and their pair potential becomes $U_a=U_b+\epsilon_b=-7\epsilon$ (in this example, $U_b=-\epsilon$ and $\epsilon_b=-6\epsilon$). If the type of at least one atom in the interacting pair is $z$, their interaction potential has an additional square shoulder of width $w_z$ and height $\epsilon_z$ [blue line, panel (c)]. Both blue and black lines go to infinity at the bond distance $w_b$, to satisfy the Hamiltonian condition.
    Panels (d,~e,~f) are the same as (a,~b,~c) but for the case of sulfur ($z=2$). Atoms interact with the square well potential of width $w>w_b>\sigma$ and energy $-\epsilon$ [black line, panel (d)] if their types are not both equal to $z$: $i<z$ and $j\leqslant z$. If they come closer than the bond width $w_b$, they change their types to $i+1$ an $j+1$ and form a bond [magenta line, panel (e)] and their pair potential becomes $U_a=U_b+\epsilon_b$ (in this example, $U_a=-\epsilon$ and $\epsilon_b=3$). If the types of both atoms in the interacting pair are $z$, their interaction potential has an additional square shoulder of width $w_z$ [blue line, panel (f)]. Both blue and black lines go to infinity at the bond distance $w_b$, to satisfy the Hamiltonian condition.} 
    \label{fig:1}
\end{figure}

The original DMD limited valence model for spherically symmetric step potentials is described in reference~\cite{Shumovsky2022}. The potential energy of the system changes when the particles collide, forming bonds or coming to the distance from each other at which the interparticle potential has a step. Moreover, certain pairs of particles when they come to a
particular distance from each other can react and change their types. Some of the reactions
are bond forming and they are reversible, so that when the bonded pair of particles come to the distance equal to the limited bond length, the bond can break and the particles change their types back to their original types before the bond formation.
The model can produce a cascade of reactions in which the type of particles is determined by how many bonds they have, so if a particle $i$ of type $k$ with $k$ bonds react with a particle $j$
with $l$ bonds they can form a new bond and their types become $k+1$ and $l+1$, respectively, 
if $k+1\leqslant z$ and $l+1\leqslant z$, where $z$ is the maximum valence. The potential energy of the system changes by  $\Delta U$, where  $\Delta U$ is the potential energy change of the system associated with the changes of the types of the reacting particles with each other and due to their interactions with all other particles. $\Delta U$ also includes a term associated to the change of the potential energy of the interacting pair due to the bond formation.  Note that if before the reaction a pair potential energy of the reacting pair is $U_b$, after the reaction the pair potential energy is $U_a=U_b+\epsilon_b$, where $\epsilon_b$ is the bond energy. In the example of figure~\ref{fig:1}(b), $\epsilon_b=-6\epsilon$. Before the reaction, the pair of reacting particles separated by the reaction distance $w_b$ has only the energy of van der Waals interaction
$U_b=-\epsilon$, while after the reaction the pair has a potential energy $U_b+\epsilon_b=-7\epsilon$.  

Conversely, the bond between the two particles of types $k>1$ and $l>1$  breaks if their distance exceeds the bond length. The particles change their types to $k-1$ and $l-1$, respectively, and the potential energy of the system changes by the bond energy $-\Delta U$. The kinetic energy of the colliding pair changes by the opposite value so that the total energy is conserved. Besides this, the colliding particles change their velocities so that the total linear and angular momenta of the pair are conserved. The laws of energy and momenta conservation give six scalar equations from which the new components
of velocities of the colliding pairs can be found by solving a quadratic equation. If this quadratic equation~\cite{Sitges} does not have real roots, the particles does not change their types,
the bond does not break or form and the particles bounce off like hard spheres by changing the signs of their velocity components in the reference frame of the center of mass of the colliding pair. 

This scheme possesses almost
all features for the implementation of the Hamiltonian formalism except that we must make sure that the particles of bond forming types cannot exist within the bond forming range without forming a bond, so that the number of bonds of each particle exactly corresponds to its type. This can be achieved by adding an effectively infinite repulsive shoulder with the diameter equal to the bond-forming distance for the interaction potential of the pair of particles with the bond forming types. We chose the value of this quasi-infinite shoulder to be $100\epsilon$, where $\epsilon$ is the energy of the van der Waals interactions. 

Another feature of the DMD algorithm that was used in references~\cite{Shumovsky2022,Shumovsky2023} is that each atom is split into the ``shell'' which is a bond forming reactive part  and the ``core'' which is inert and is responsible for the van der Waals long-range interactions. The shell and the core were connected by a short infinite square well bond of length $d$. In this work, we eliminate that unnecessary split and assign both the van der Waals interactions and bond formation ability to a single particle representing an entire atom, which significantly increases
the computational efficiency of the model, because it reduces the number of particles by a factor of two.
We verify that the phase diagram  of the core-shell model converges to the phase diagram of the
single-particle model when $d\to 0$.

In principle, the limited valence model allows us to construct the tables of interatomic potentials $U_{ij}(r)$ and bond potentials $U^b_{ij}(r)$ for bonds between atoms of different types as complex as we wish, with multiple
steps at different distances $r$. This might be necessary in order to accurately reproduce the behavior of various compounds. For example, one may want to make interatomic potentials consisting of many small steps to create a ramp like in the Jagla model~\cite{Xu2005,Xu2006} or make a bond potential resembling a parabola of a harmonic oscillator~\cite{Gang2016}. But in this paper our goal is to keep the model as simple as possible with potentials being a square well potential or a combination of a square well and a square shoulder. Moreover, we choose interatomic potential to be the same for atoms
of all types $i<z$ and $j<z$, adding to it a repulsive shoulder if $i=z$ or
$j=z$ to model dissociation of bonds under pressure as in hydrogen, or adding an attractive well if both $i=z$ and $j=z$ to model association under pressure as in sulfur. This reduces the number of parameters of the model to
six: the maximum valence $z$, the width of van der Waals attraction well, $w$, (the diameter of hard core of this well, $\sigma$, and the depth of this well
$\epsilon$ serving, respectively, as units of length and energy), the width of the bond $w_b$, the energy of the bond, $\epsilon_b$, the width of the repulsive shoulder or an attractive well, $w_z$, and the height of the shoulder (depth of the well),  $\epsilon_z$, with $\epsilon_z>0$ for repulsion and $\epsilon_z<0$ for attraction.  
 Figure~\ref{fig:1}(a,~b,~c) illustrates the potentials of the single-particle model for the case when the atoms with maximum valence have an additional potential shoulder which repel all types of atoms [figure~\ref{fig:1}(c)].  This variant is used to model hydrogen and water.  Figure~\ref{fig:1}(d,~e,~f) illustrates the potentials for the case when the atoms with maximal valence have an additional attractive well which attracts only atoms of maximum valence  [figure~\ref{fig:1}(f)]. This variant is used to model sulfur.

To summarize, the model has three essential potentials, without which it cannot reproduce
the phenomenon of liquid polyamorphism for all values of $z>0$:

\noindent the van der Waals potential:
\begin{equation}
U_{ij}(r)=\left\{\begin{array}{lll}
\infty && r\leqslant w_b,\\
-\epsilon &&w_b<r\leqslant w,\\
0 && r>w,
\end{array}
\right.
\label{eq:vdw1}
\end{equation}
the bond potential for $0<i\leqslant z$ and $0<j\leqslant z$:
\begin{equation}
U_{ij}^b(r)=\left\{\begin{array}{lll}
\infty&& r\leqslant \sigma,\\
\epsilon_b-\epsilon &&\sigma < r\leqslant w_b,\\
0 &&r>w_b,
\end{array}
\right.
\label{eq:bond}
\end{equation}
and the modified van der Waals potential:
\begin{equation}
U_{ij}(r)=\left\{\begin{array}{lll}
\infty&& r\leqslant w_b,\\
\epsilon_z-\epsilon && w_b < r\leqslant w_z,\\
-\epsilon && w_z <r\leqslant w,\\
0 && w<r.
\end{array}
\right.
\label{eq:vdw2}
\end{equation}
Note, that matrix $U_{ij}$ is symmetric. For the case of dissociation under pressure ($\epsilon_z>0$ like in hydrogen and water), we take
 $i < z$ and $j<z$ in equation  (\ref{eq:vdw1}) and we take
$j=z$ or $i=z$ in  equation  (\ref{eq:vdw2}). In the case of association under pressure ($\epsilon_z <0$ like in sulfur) we take $i<z$ and $j\leqslant z$  in equation (\ref{eq:vdw1}) and we take $i=j=z$ in equation (\ref{eq:vdw2}). In principle, we can take $\epsilon=0$ if $z\geqslant 3$
because the liquid for $z\geqslant 3$ can form due to multipe bonds between the particles
wihout van der Waals attraction~\cite{Zaccarelli2005}.

All results below are presented in dimensionless units based on the three basic units~\cite{Shumovsky2022, Shumovsky2023}: $\sigma$, the hard core diameter is the unit of length; $\epsilon$, the depth of the van der Waals potential well is
the unit of energy; and the mass of a particle, $m$,  is the unit of mass. We perform all simulations in the NVT ensemble with a constant number of atoms
$N=1000$, in a cubic box with periodic boundaries of fixed volume $V$ such that the density $\rho=N/V$ has a given value. The temperature is controlled by the Berendsen thermostat~\cite{Berendsen1984} modified for the DMD algorithm~\cite{Sitges}. To estimate the location of the critical points, we first perform preliminary simulations of the single particle model for $10^5$ time units by slowly cooling the system from the initial temperature greater than critical temperature to a very low temperature which is below all interesting features of the phase diagram, saving average temperature, pressure, and potential energy every 1000 time units, which is well above the equillibration time, estimated to be below $500$ time units in the entire range of temperatures and densities studied. Then, for selected points in the parameter space we perform long equillibration runs of $10^5$ time units for each state point.

\section{Liquid--liquid phase transition associated with dimerization ($z=1$)}
\label{sec:z=1}
\begin{figure}[t]
   \centering
    \includegraphics[width=0.49\linewidth]{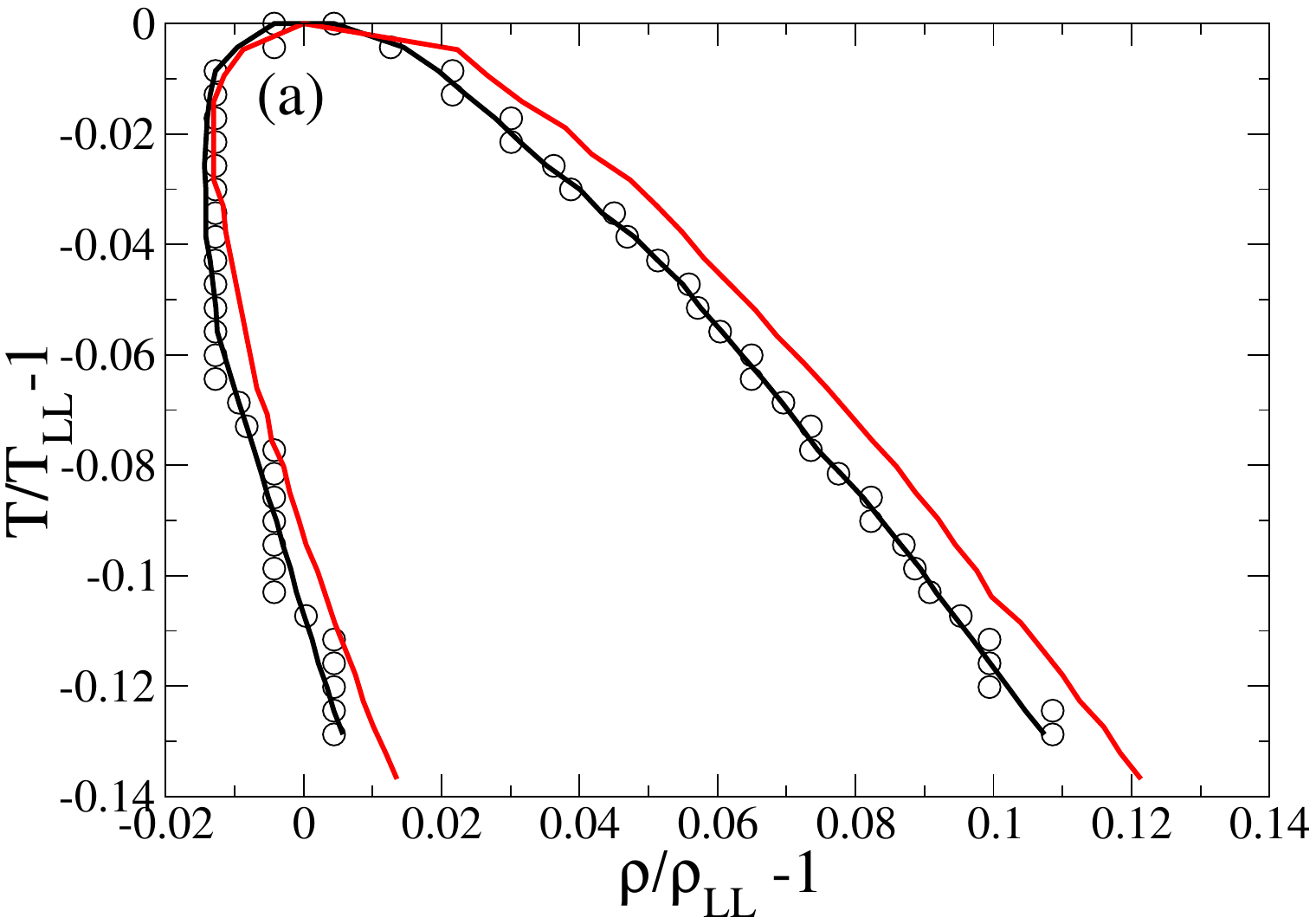}
    \includegraphics[width=0.49\linewidth]{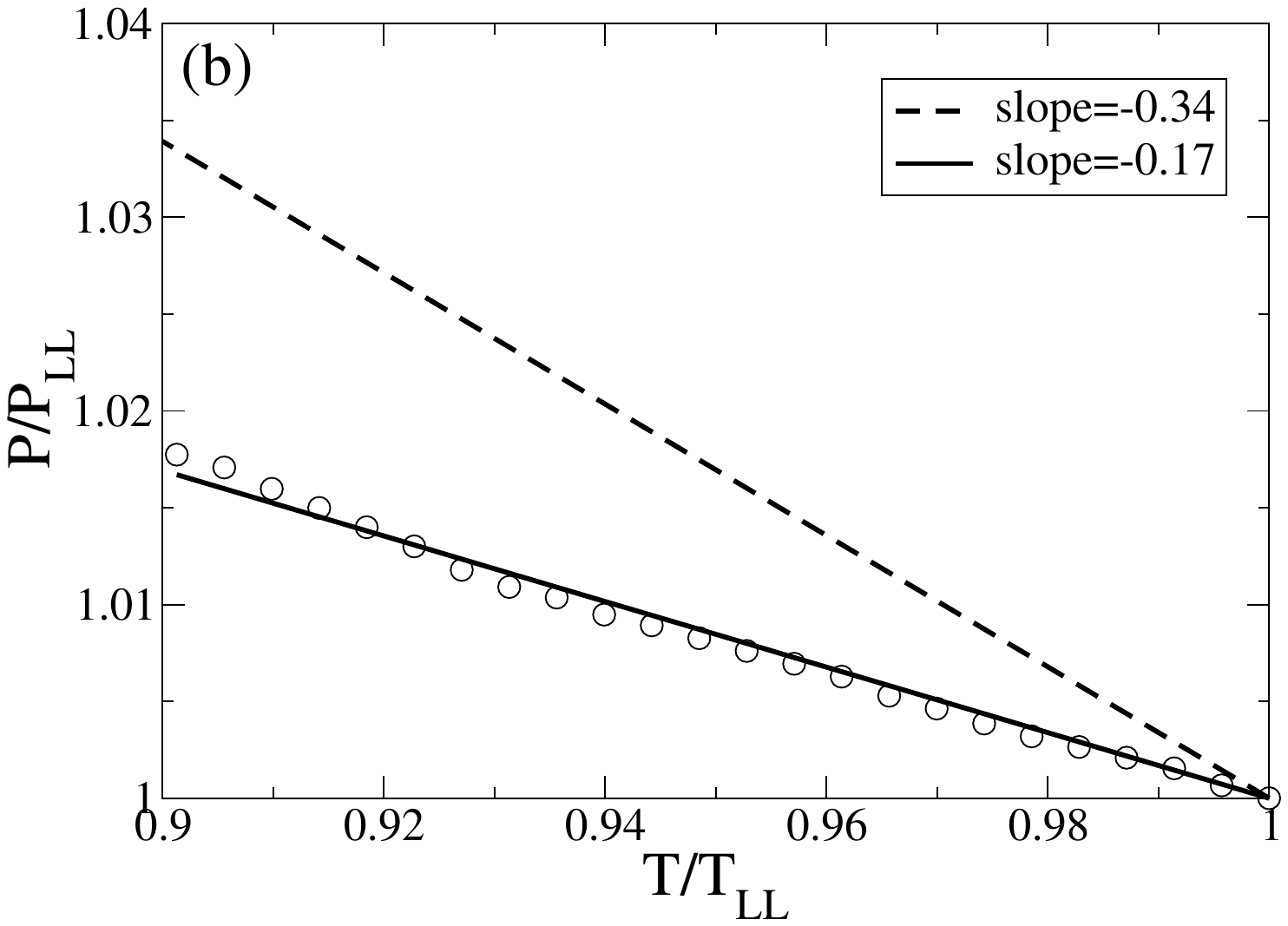}
    \caption{(Colour online) Comparison of the liquid--liquid coexistence lines  of the single-particle Hamiltonian maximum valence  model and the core-shell non-Hamiltonian model studied in reference~\cite{Shumovsky2023} for  $z=1$, $w=1.5$, $w_b=1.1$, $w_z=1.2$, 
    $\epsilon_b=-6$, $\epsilon_z=12$. (a) Coexistence line in the $\rho-T$ plane in reduced coordinates.  Circles are the results obtained by Maxwell construction for the Hamiltonian model. The jumps in the results are due to the discreteness of the values of $\rho$ which are produced with step $\Delta\rho =0.005$.   The black curve shows the same results after a smoothing procedure. The red curve shows the results for the core-shell model without the Hamiltonian restriction. One can see that the results are very close to each other, but the Hamiltonian variant produces a slightly narrower coexisting curve but not significant in reproducing the hypothesized transition for real hydrogen~\cite{Fried_Hydrogen_2022}. (b) The coexisting lines on the $P-T$ plane. The dashed straight line is the straight line fit of the results of reference~\cite{Shumovsky2023}. One can see that in reduced
    coordinates, the slope of the coexistence line in the Hamiltonian model is two times smaller than in the non-Hamiltonian 
    core-shell model, However, it is still negative as it is predicted for real hydrogen~\cite{Fried_Hydrogen_2022}.
   } 
    \label{fig:2}
\end{figure}

\begin{figure}[t]
   \centering
    \includegraphics[width=0.32\linewidth]{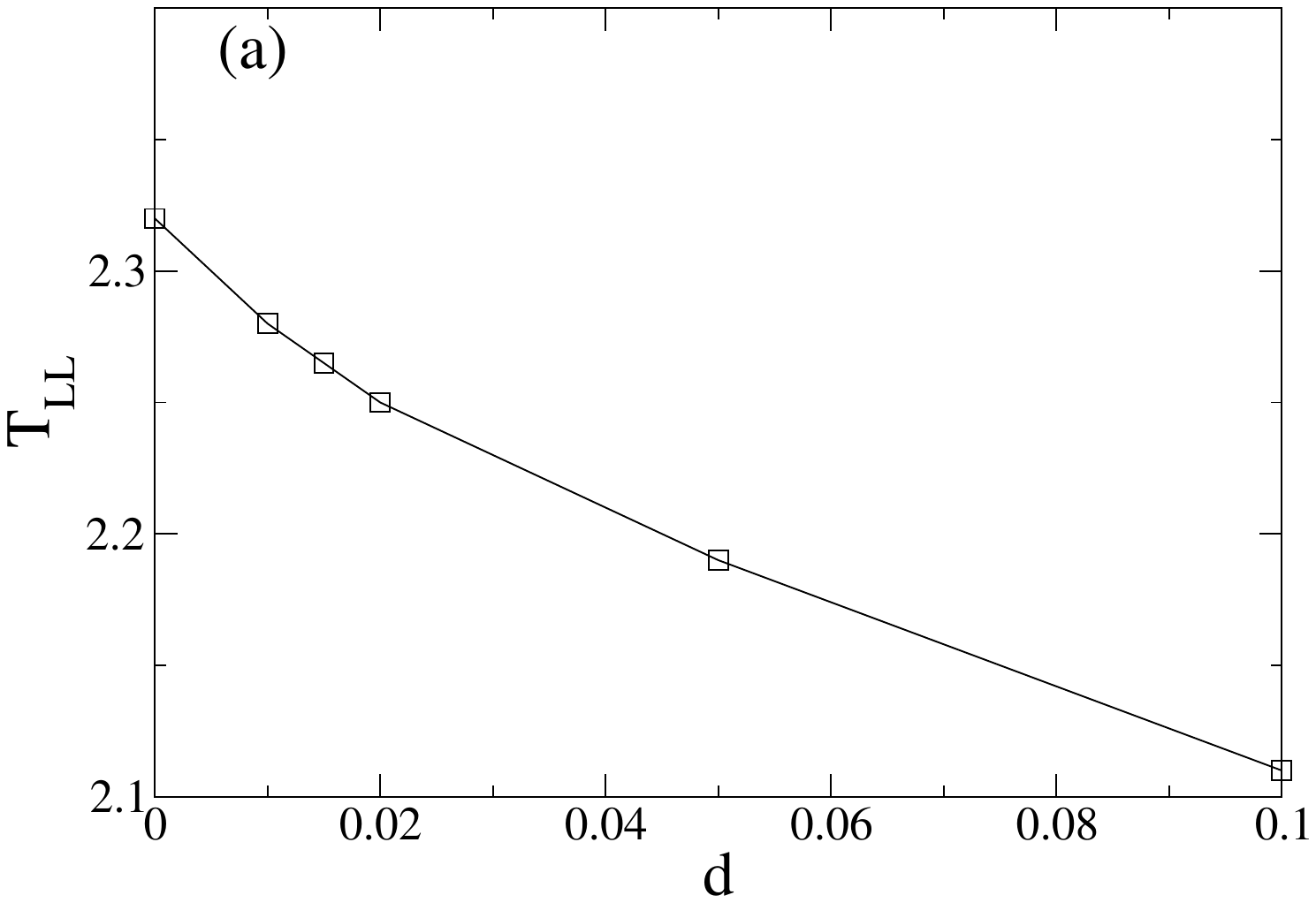}
    \includegraphics[width=0.32\linewidth]{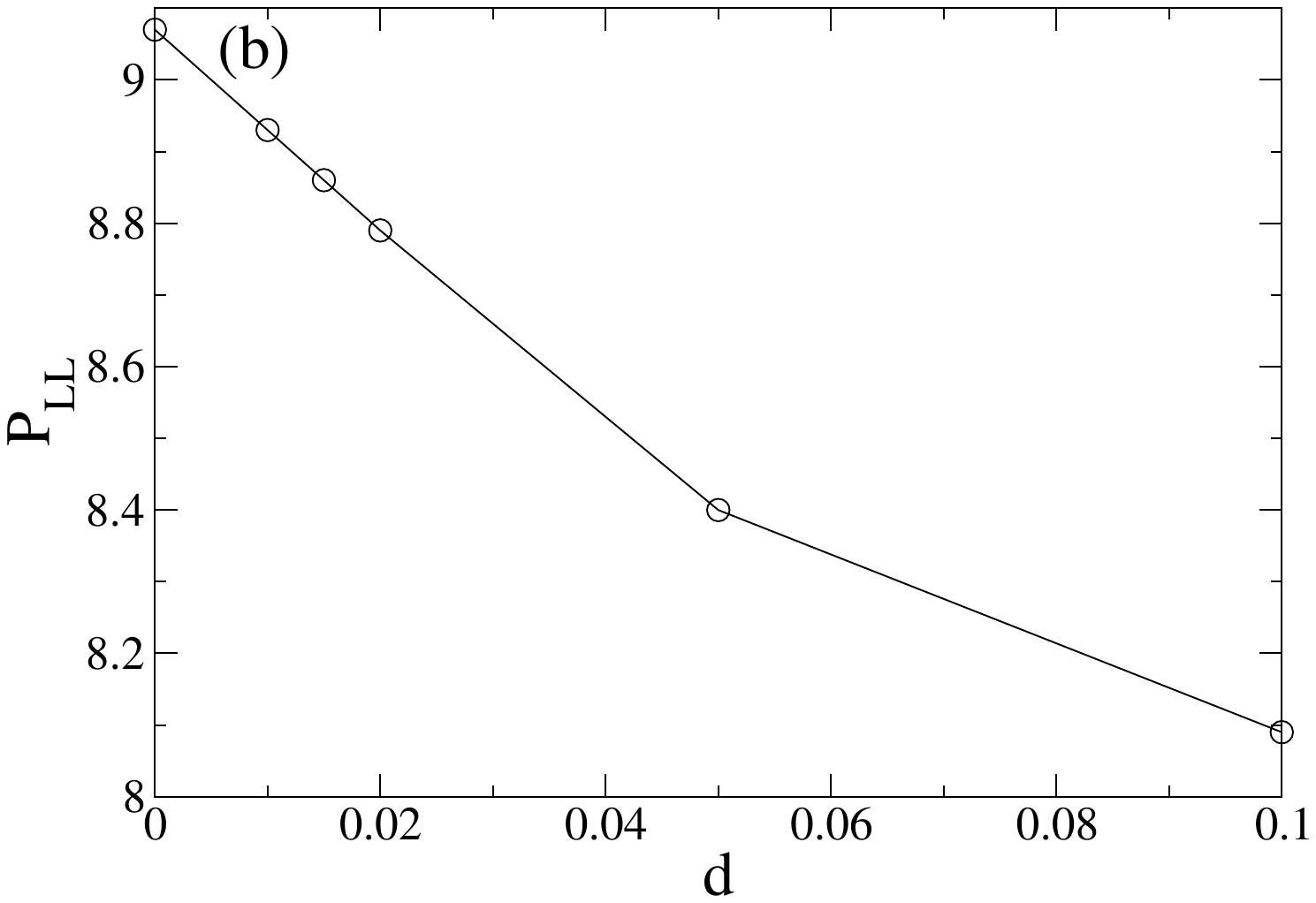}
    \includegraphics[width=0.32\linewidth]{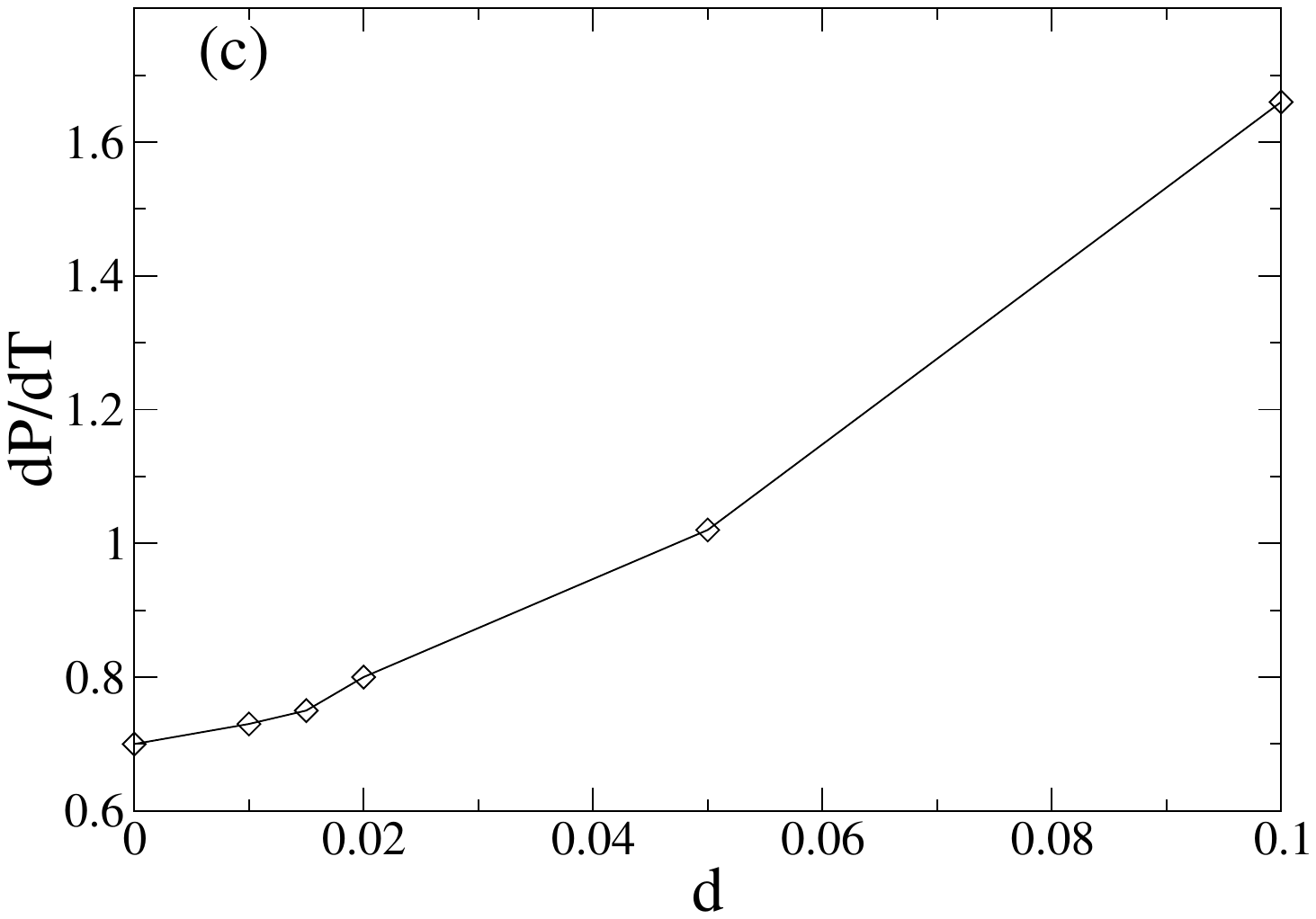}
    \caption{Convergence of the phase diagram of the core-shell model to that of the single-particle model as the bond length connecting the core and the shell tends to zero. The parameters of the model are taken for the case attempting to mimic dimerization of metallic hydrogen: $z=1$, $w=1.5$, $w_b=1.1$, $w_z=1.2$, 
    $\epsilon_b=-6$, $\epsilon_z=12$. Panels (a,~b,~c) show the parameters of the critical points $T_{LL}$, $P_{LL}$ and $dP/dT$ as functions of $d$. One can see that their
    dependence on $d$ is almost linear, including the last point for the single-particle model $d=0$.
    } 
    \label{fig:3}
\end{figure}

The results for the single-particle model with the Hamiltonian restriction does not differ much from the previously obtained ones for the core-shell model without the Hamiltonian restriction~\cite{Shumovsky2023}.  Here, we tested the case of $w_z=1.1$, $w_b=1.2$, $\epsilon_z=-6$, $\epsilon_b=12$, which corresponds to almost totally impenetrable shoulder.
We find the LLCP at $T_{LL}=2.33$, $P_{LL}=9.07$, $\rho_{LL}=0.575$ for the single-particle Hamiltonian model instead of the previously found~\cite{Shumovsky2023} for 
the core-shell model $T_{LL}=2.12$, $P_{LL}=8.17$ and the same density $\rho_{LL}$.
The shape of the coexistence curve on the $\rho-T$ plane in reduced coordinates closely follows the tilted parabola published in figure~\ref{fig:3} of reference~\cite{Shumovsky2023} [figure~\ref{fig:2}(a)].
We also carefully tested the slope of the coexistence line on the $P-T$ phase diagram [figure~\ref{fig:2}(b)] and found a definitely negative slope $\rd p/\rd T=-0.7\pm 0.2$, but this value is two times smaller than for the maximum valence core-shell model without the Hamiltonian restriction. 

To understand what causes this difference, we perform the simulations
of the core-shell model with the Hamiltonian restriction and compare it with the results of the same model without the Hamiltonian restriction
as in reference~\cite{Shumovsky2023}. We obtained very similar results: $\rho_{LL}=0.575$, $T_{LL}=2.11$, $P_{LL}=8.07$, and $\rd P/\rd T=-1.6$.
This is not surprising, because the number of shells which stay within 
the bond distance from another shell without forming a bond and  without the potential energy loss associated with the bond is small in
the range of temperatures and densities we studied. For example, for $T=2.11$, $\rho=0.58$, in the bond range of the shell of type 1, there are on average 0.015 non-bonded intruders, while in the bond range of the shell of type 0, there are on average 0.005 intruders. Such small values are caused by a very high repulsive shoulder $\epsilon_z=12$.
For smaller values of $\epsilon_z$, we expect
stronger differences between the Hamiltonian and the non-Hamiltonian variants of the model.

The larger discrepancy with the Hamiltonian single-particle model is caused by the fact
that the core and the shell of the same atom do not exactly coincide with each other but
are connected by the permanent bond of distance $d=0.1$. In principle,
the Hamiltonian core-shell model must converge to the Hamiltonian single-particle model when $d\to 0$. We test this convergence in figure~\ref{fig:2}. The values of $\rho_{LL}$ remain $0.575$ for the entire
range of $d$, while the values of $T_{LL}$, $P_{LL}$ and $\rd P/\rd T$ change
almost linearly with $d$, smoothly converging to the values of the single-particle model ($d=0$).

\begin{figure}[t]
    \centering
    \includegraphics[width=0.49\linewidth]{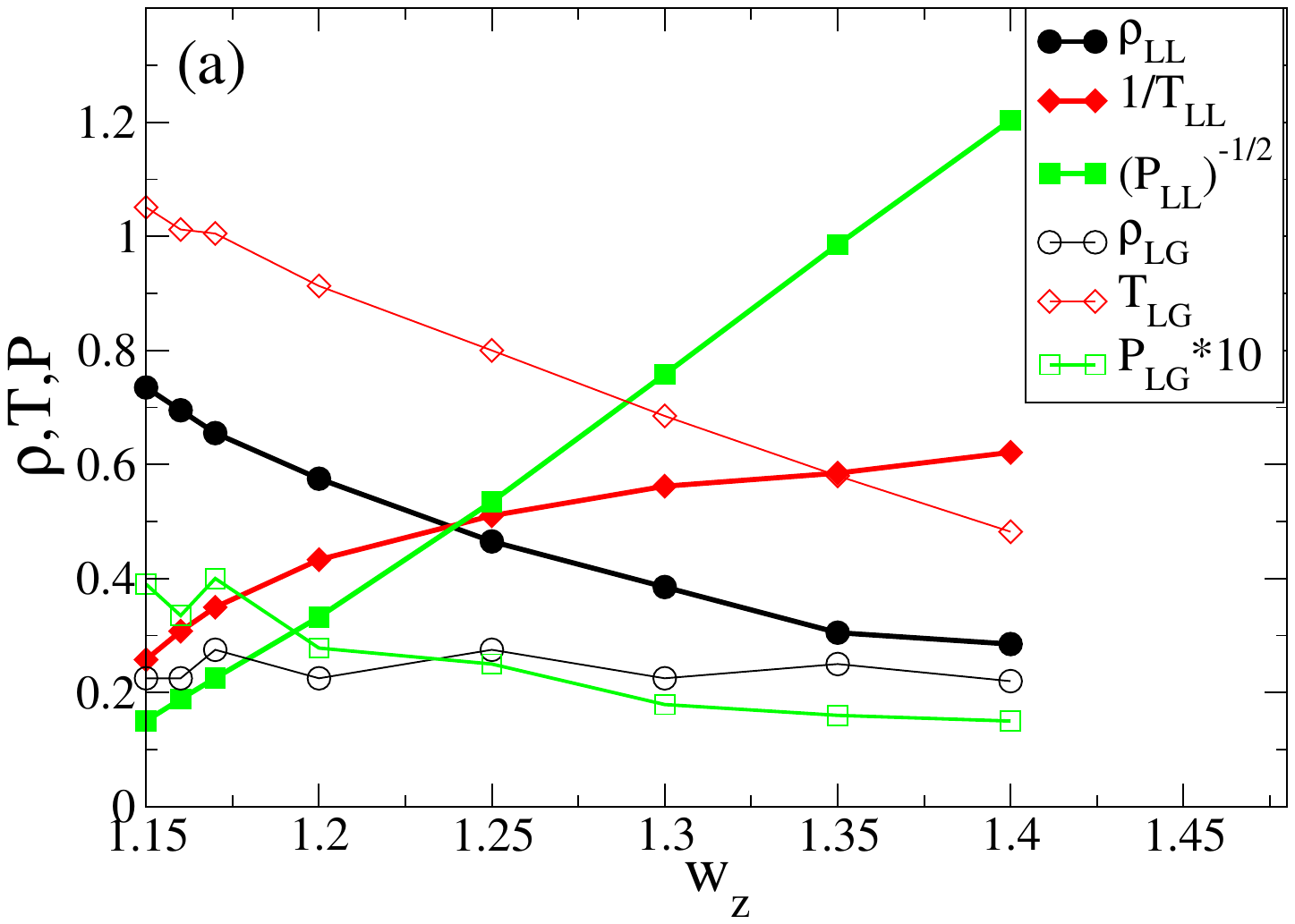}
     \includegraphics[width=0.49\linewidth]{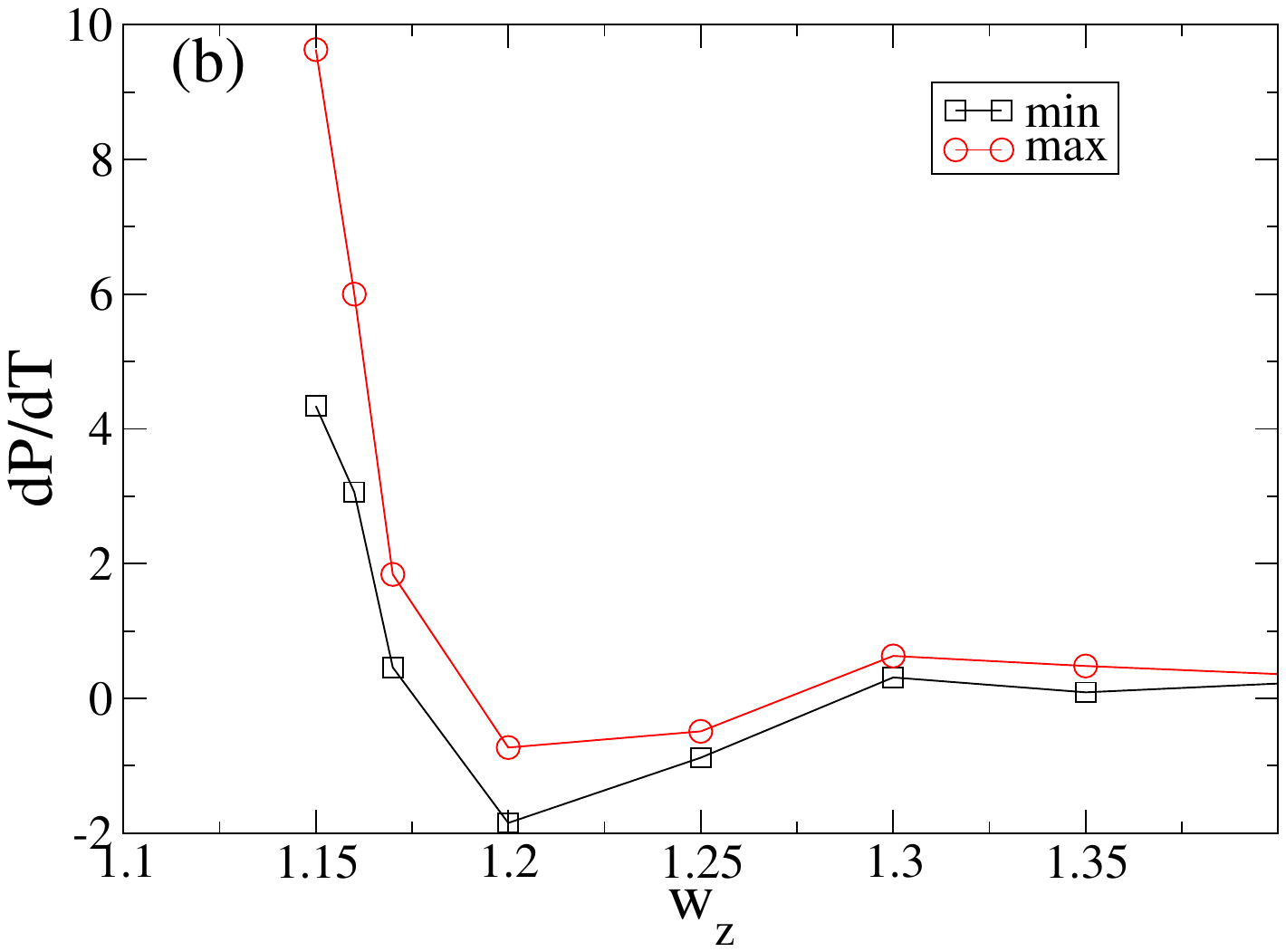}
    \caption{(Colour online) (a) Dependence of the location of the critical points on the repulsive shoulder width, $w_z$  for the dimerization model $z=1$, $w=1.5$, $w_b=1.1$, $\epsilon_b=-6$ and $\epsilon_z=13$. The critical density decreases from 0.735 at $w_z=1.15$ to $0.285$  at $w_z=1.4$. The critical temperature decreases by factor of 3 and pressure decreases even more dramatically, approximately as $0.056(w_z-w_b)^{-2}$. (b) The slope of the coexistence line
    versus $w_z$ estimated by the slopes of the two isochores with step $\delta\rho=0.01$ crossing in the $P-T$ plane at the maximum temperature, which is also how we obtain the location of the critical points in panel (a).
    } 
    \label{fig:3b}
\end{figure}

In addition, we study the effect of the width of the repulsive shoulder $w_z$ on the location of the critical points and the slope of the coexistence line (figure~\ref{fig:3b}). We observe that the critical density, temperature, and pressure of the LLPT all strongly decrease when the width of the repulsive shoulder increases. By contrast, the slope of the coexistence line behaves non-monotonously with a minimum at $w_z=1.2$, where $\rd P/\rd t$ is definitely negative as hypothesized to be the case for real hydrogen~\cite{Fried_Hydrogen_2022}.

At the widest repulsive shoulder $w_z=1.4$, the dimeric (molecular) liquid becomes metastable with respect to the monomeric (metallic) liquid. This is because the van der Waals effective attractive  range of dimers becomes just $w-w_z=1.5-1.4=0.1$, compared to practically impenetrable shoulder of $w_z=1.4$. Such a system would have a very low temperature  of the liquid--gas critical point (LGCP) which is metastable to crystallization~\cite{Xu2012}. 
However, the dimeric liquid also becomes metastable with respect to the metallic liquid in which monomers lacking the repulsive shoulder may have 12 or more neighbors in the
wide van der Waals well of $w=1.5$, which gives the average potential energy per particle $-6\epsilon=-6$. If we assume that an atom in the dimeric liquid has 11 neighbors in a very thin attractive well (one place is taken by the atom forming a bond with energy $-6$), the potential energy per atom is $(-11 -6)/2=-8.5$, but the entropy cost for forming such a dense configuration is such that the liquid near the critical point never achieves such a density and its free energy is greater than that of monomeric liquid.

\begin{figure}[t]
    \centering
    \includegraphics[width=0.5\linewidth]{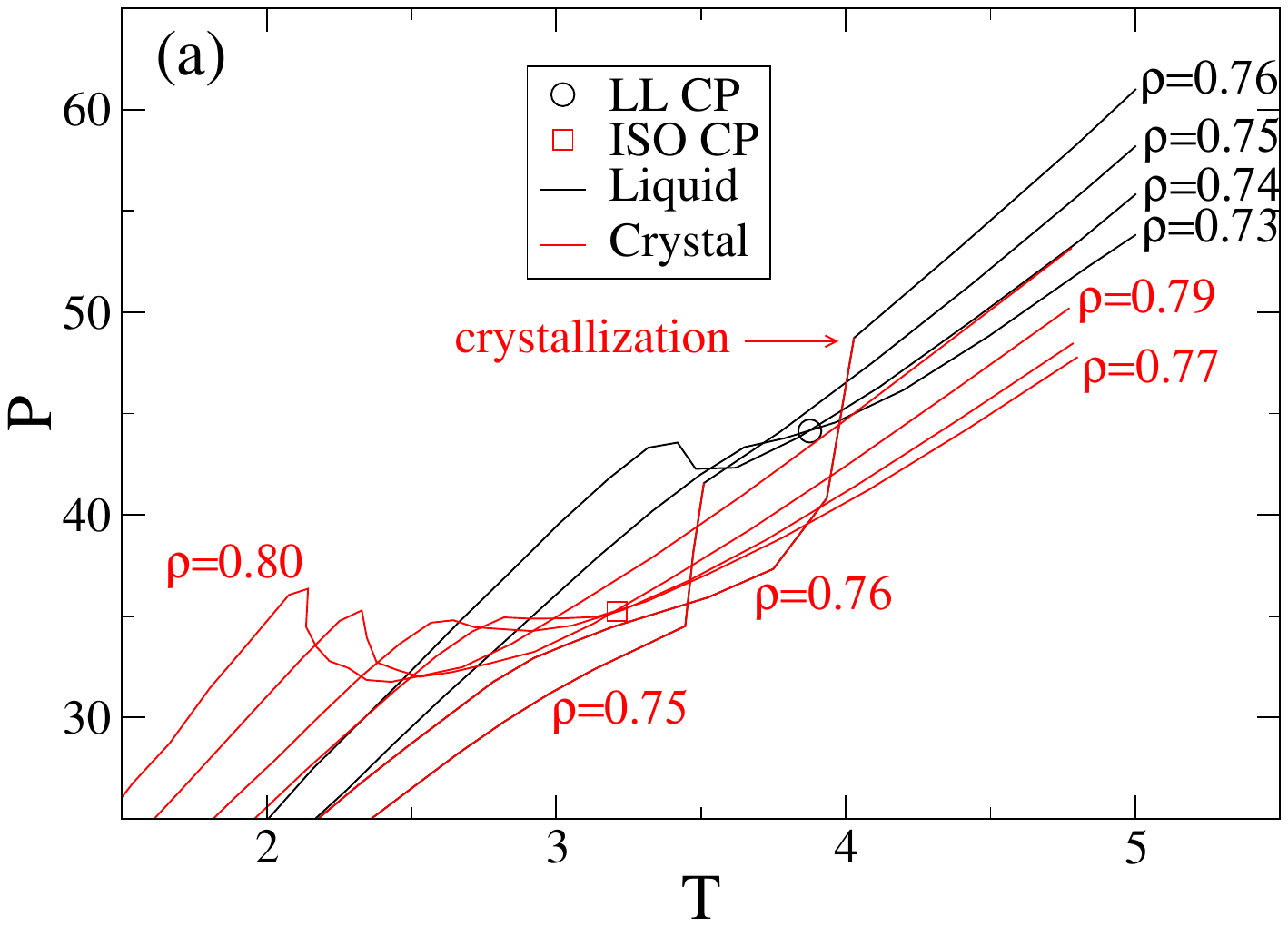}
     \includegraphics[width=0.4\linewidth]{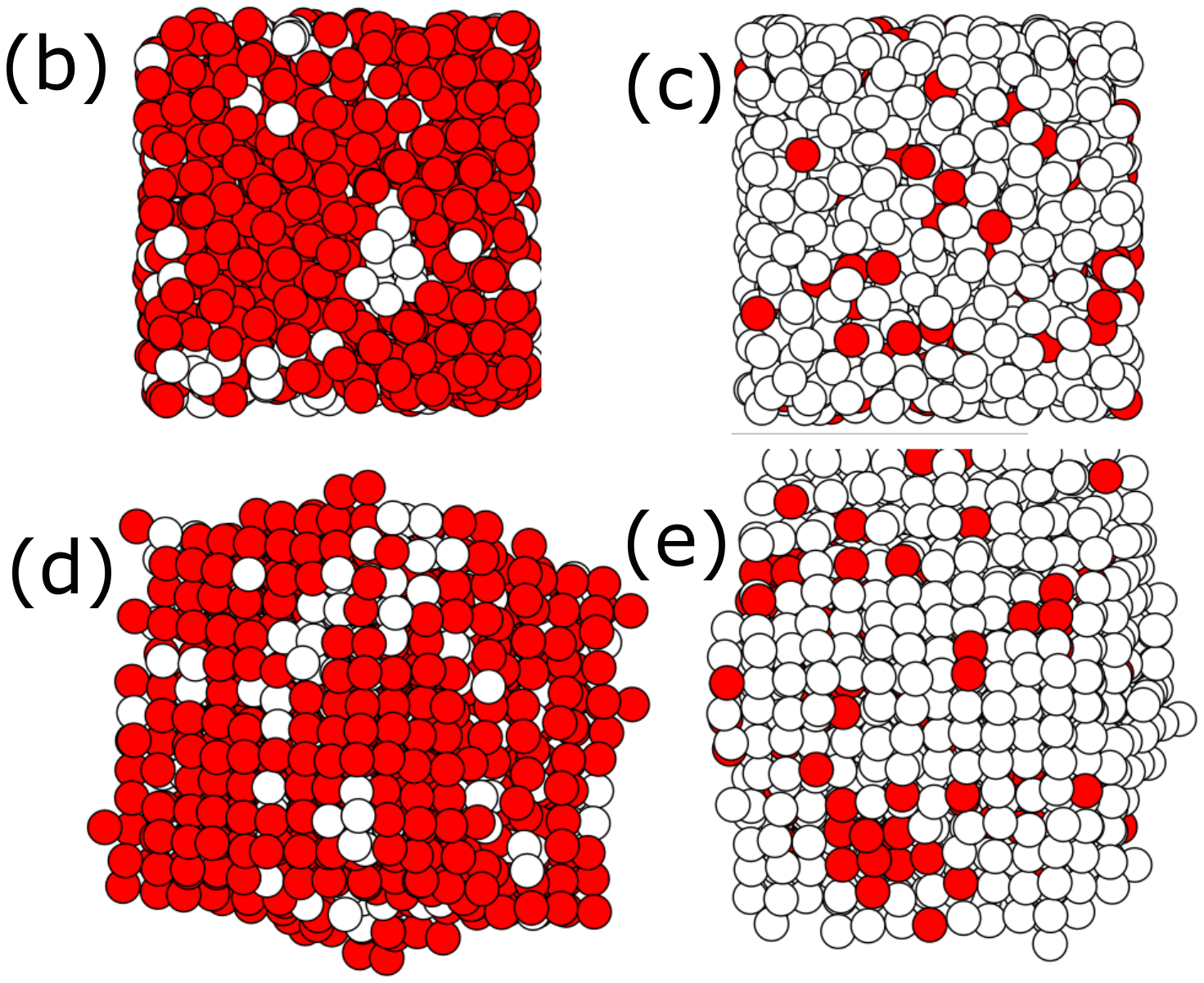}
    \caption{(Colour online) (a) The $P-T$ phase diagram of the dimerization model ($z=1$) with $w_z=1.15$,  $w=1.5$, $w_b=1.1$, $\epsilon_b=-6$ and $\epsilon_z=13$. The black
    curves are parts of the isochores corresponding to the liquid state, and the red curves are parts of the isochores corresponding to crystalline state, obtained by slow cooling with equilibrarion time of 1000 time units. The crystallization event is shown at the isochore with $\rho=0.77$. The two critical points are visible: the liquid--liquid one indicated by a circle and the isostructural solid--solid  one indicated by a square. (b) The snapshot of the system taken along the the liquid--liquid critical isochore $\rho=0.74$ above the critical point at $T=4.44$, $P=49.7$, (c) the snapshot of the system taken along the same isochore $\rho=0.74$ below the critical point at $T=3.02$, $P=39.5$, (d) the snapshot of the system taken along the solid--solid critical isochore $\rho=0.77$ above the critical point at $T=4.8$, $P=47.7$, (e)  the snapshot of the system taken along the the solid--solid critical isochore $\rho=0.77$ below the critical point at $T=2.42$, $P=31.4$. In (d) and (e) one can clearly see the crystalline plane (1,0,0) of the fcc lattice. Atoms of type 0 (monomers) are shown in red. Atoms
    of type 1 (belonging to dimers) are shown in white.
    } 
    \label{fig:4}
\end{figure}

At the narrowest repulsive shoulder $w_z=1.15$, we observe spontaneous crystallization into the face centered cubic (fcc) crystal upon cooling for $\rho\geqslant 0.75$, indicated by the abrupt drop of pressure, because the ordered fcc crystal occupies less volume than the disordered liquid. The LLPT is still present at $\rho=0.735$, $T=3.88$, $P=0.441$. In addition, we observe the isostructural phase transition between two fcc crystals with different lattice constants. This phase transition ends at the isostructural critical point~\cite{Kincaid1976} at larger $\rho=0.78$, but smaller $T=3.21$ (figure~\ref{fig:4}). It is worth mentioning that the isostructural critical point was discovered experimentally and was explained theoretically using core-softened potentials~\cite{Kincaid1976,Frenkel94} long before the LLCP~\cite{Franzese2001}. The systems with core-softened potentials with a narrow repulsive shoulder are prone to crystallization at high density, while the maximum valence model
based on the same potentials, with the restriction that only $z$ neighbors can
enter the shoulder, is not~\cite{Smallenburg2013}. Here, we observe a novel situation, when both the solid--solid and the liquid--liquid critical points can be observed simultaneously.  

\section{Liquid--liquid phase transition in tetrahedral networks ($z=4$)}
\label{sec:z=4}
The hypothetical LLPT in the supercooled water is associated with the reordering of the tetrahedral network of hydrogen bonds~\cite{Gallo_Water_2016}. The justification of the application of the limited valence model with $z=4$ is as follows. In the low density liquid, each oxygen is linked with its four neighbors by 4 hydrogen bonds forming a rigid first coordination shell, penetration into which has a high energy cost modelled by a repulsive shoulder $w_z$. 
These oxygens are assumed to be of type 4. In the high density liquid, some of the bonds are bifurcated. These bonds have much higher potential energy than the normal bonds~\cite{Giguere1984}. Thus, in our model we assume that the oxygens which have a bifurcated bond have only 3 bonds and these oxygens belong to type 3. They are not surrounded by a repulsive shoulder and hence they allow other oxygens of type 3 to enter their first
coordination shell. The core-shell variant of this model has been already studied in reference~\cite{Shumovsky2023} and shows the LLPT with a negative slope of the coexistence line, surrounded by a region of density anomaly on the $P-T$ plane. The region of density anomaly existed only at very high pressures and did
not reach $P=0$ line like in real water.

\begin{figure}[t]
    \centering
    \includegraphics[width=0.75\linewidth]{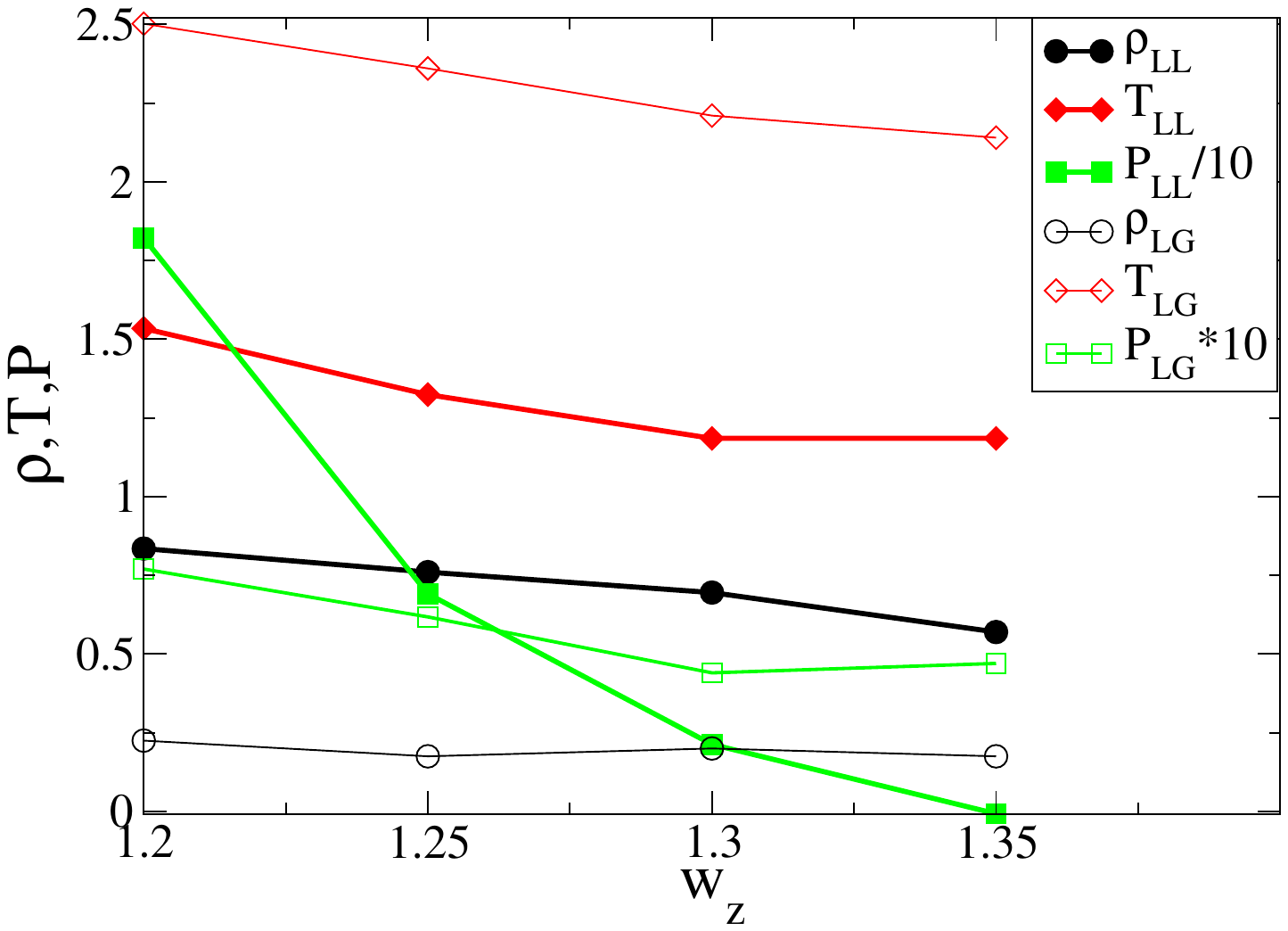}
    \caption{(Colour online) The dependence of the location of the critical points for the tetrahedral model $z=4$, $w=1.5$, $w_b=1.1$, $\epsilon_b=-6$, $\epsilon_z=6$  on the width of the repulsive shoulder $w_z$. One can see that as $w_z$ increases, the critical density, temperature and pressure decrease, until the pressure becomes negative and the LLCP disappears below the liquid gas spinodal for $w_z\geqslant 1.4$. The liquid--gas
    critical temperature and pressure also decreases. The liquid--liquid critical temperature is always below the liquid--gas critical temperature as in water.
    }   
    \label{fig:5}
\end{figure}

\begin{figure}[t]
    \centering
    \includegraphics[width=0.49\linewidth]{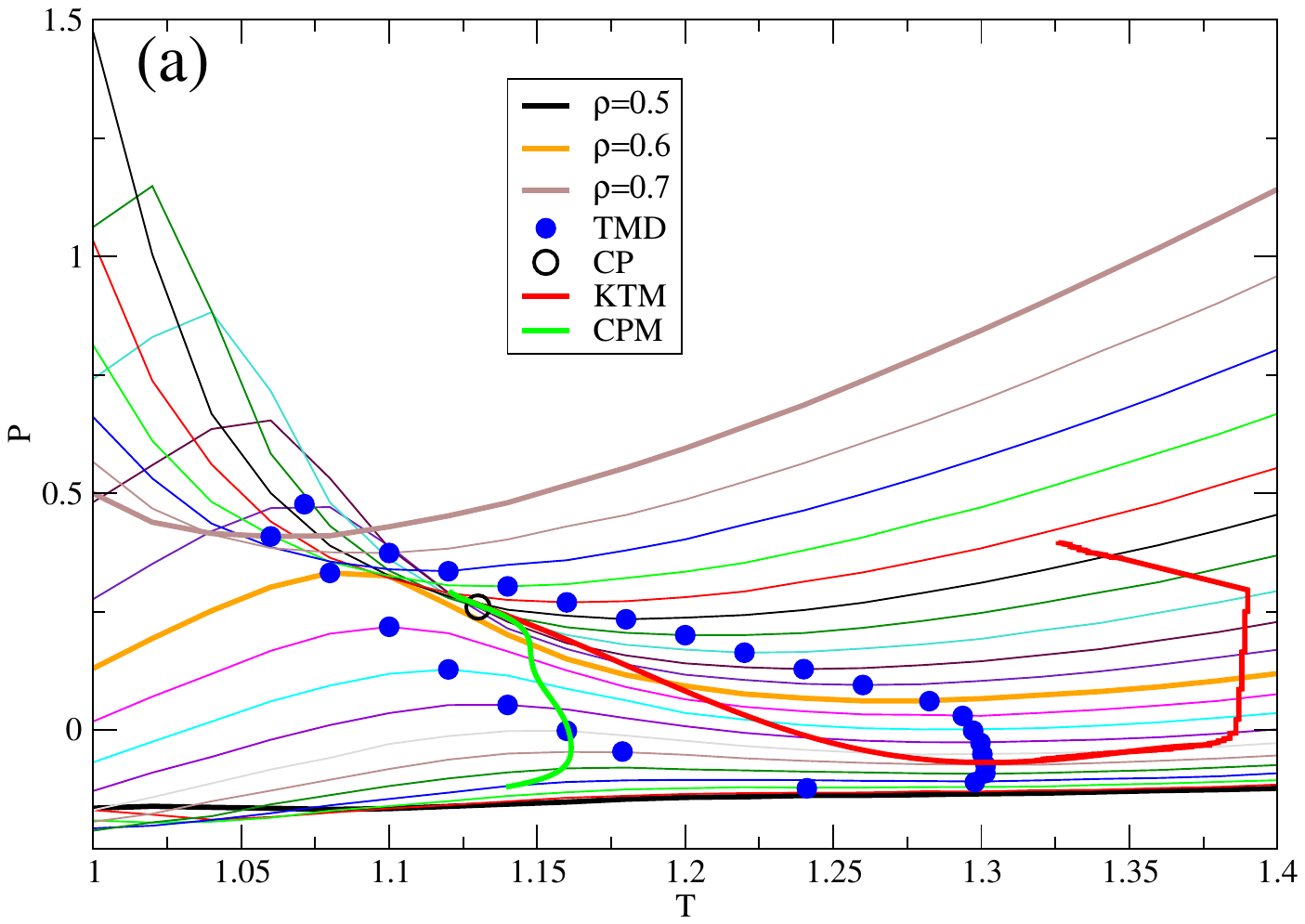}
     \includegraphics[width=0.49\linewidth]{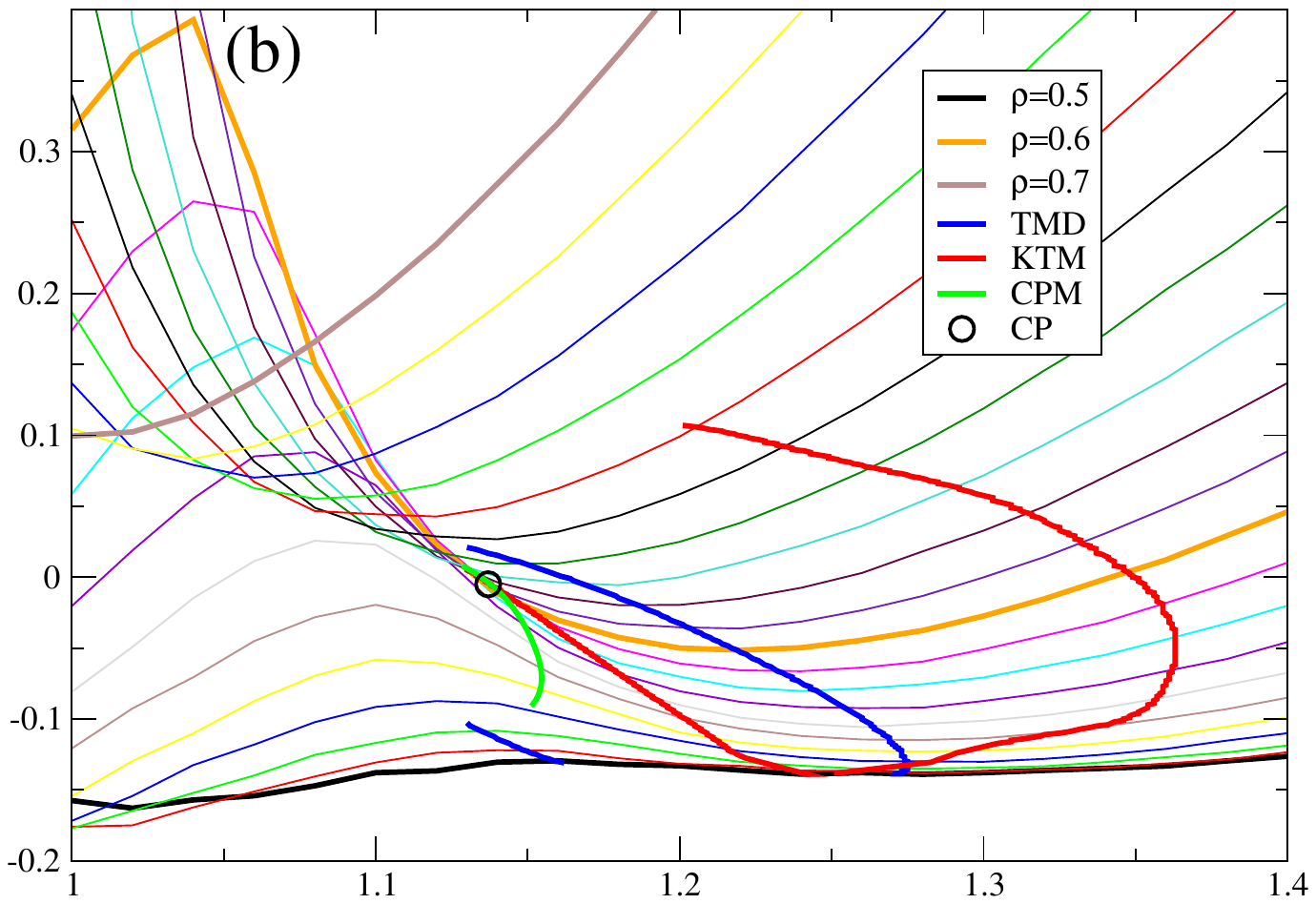}
    \caption{(Colour online) $P-T$ phase diagrams of the $z=4$ with  $w=1.5$, $w_b=1.1$, $\epsilon_b=-6$ and $\epsilon_z=6$ for $w_z=1.34$ (a) and $w_z=1.35$ (b).
    One can see the isochores crossing at the critical points
    surrounded by the region of the density anomaly as in reference~\cite{Poole2005} for the ST2 model of water. 
    This region is separated by the TMD (temperature of maximum and minimum density) line connecting the points of minimum and maximum pressure at constant density. Also
    are shown the line of isothermal compressibility, $\kappa_T$ maxima and minima (KTM) which crosses the TMD line at its vertical point~\cite{Sastry1996} and the constant pressure specific heat maxima (CPM) line (green) which coincides with the KTM line near the critical point~\cite{Sastry1996}. The TMD, KTM and CPM lines are obtained by the polynomial interpolation of the computed
    points on the isocores, obtained with steps $\Delta \rho=0.01$ and $\Delta T=0.01$. Some of the inflections and irregularities on these lines can be due to the boundaries of the fitted region.
    The low density isochores in (b) start to intersect with each other forming a liquid--gas spinodal, which becomes non-monotonious
    following the isochores with a minimum at the point of maximum density like in the IAPWS95~\cite{Wagner99}  equation of state of water extrapolated to the metastable region of negative pressures.}
    \label{fig:6}
\end{figure}

Here we will study how the change of the repulsive shoulder width $w_z$ affects the position of the LLCP
and the entire phase diagram for $z=4$. The motivation for this study is to try to relate the known experimental properties of
the water radial oxygen-oxygen distribution function $g_{\rm OO}(r)$ (see, e.g., figure~\ref{fig:5} of reference~\cite{Gallo_Water_2016}) to the shoulder width, $w_z$. 
One can see that both high density liquid (HDL) and low density
liquid (LDL) have a sharp peak of the radial distribution function at $r\approx 0.27$~nm. By contrast, the LDL have a minimum at $r=0.33$~nm absent in the HDL which instead has a shoulder at this point and meets the LDL graph at $r=3.9$~nm. In terms of the limited valence model, this means that the repulsive shoulder $w_z$ of the atoms with $z=4$ bonds (which prevents the fifth intruder to enter the vicinity of the first coordination shell) extends
up to $r\approx 3.9$~nm. Thus, we can conclude that in order to achieve good agreement with water we need to set $3.3/2.7\approx 1.2 <w_z <3.9/2.7 \approx 1.4$. 
Figure \ref{fig:5} shows the dependence of the location of the critical points on $w_z$
for $z=4$, $w=1.5$, $w_b=1.1$, $\epsilon_b=-6$, $\epsilon_z=6$. The behavior
of all the variables qualitatively resembles
their behavior for $z=1$. Except that the dependence is weaker and the critical pressure
becomes negative for the shoulder width $w_z=1.35$. The slope of the liquid--liquid coexistence line is always negative.
For $w_z=1.4$, the LLCP becomes submerged below the liquid--gas spinodal. For $w_z=1.15$,  the system
crystallizes into a body centered cubic lattice at high densities. The isostructural solid--solid critical point may exist
at even higher densities, but this question requires further investigation. Figure \ref{fig:6} shows the $P-T$ phase diagrams which resemble those of water~\cite{Gallo_Water_2016}, showing the region of density anomaly at negative pressures~\cite{Caupin2006}, the compressibility minimum and the heat capacity maximum near the critical point. 

The exact location of the hypothetical (LLCP) of water is unknown. 
The simulations of the ST2 model~\cite{Poole2005} give the value of
$P_{LL}=195$~MPa and $T_{LL}=245$~K. More recent estimates based on the TIP4P/ice model~\cite{Espinosa2023} suggest the values $P_{LL}=125$~MPa, $T_{LL}=195$~K. Theoretically, the LLCP can be at negative pressures or can even disappear below the liquid--gas spinodal~\cite{Gallo_Water_2016}. The LGCP of water is located at $P_{LG}=22$~MPa which is several times smaller than $P_{LL}$, and $T_{LG}=647$~K, which is several times larger than $T_{LL}$.
It is clear that the maximum valence model with $z=4$ can reproduce these values after some additional tweaking of parameters.
Beside this, the maximum valence model qualitatively correctly describes the water anomalies. The best agreement with water is reached at $w_z=1.34$
which does not contradict our estimates based on $g_{\rm OO}(r)$.

\section{Liquid--liquid phase transition associated with polymerization of dimers ($z=2$)}
\label{sec:z=2}
\begin{figure}[!h]
    \centering
    \includegraphics[width=0.7\linewidth]{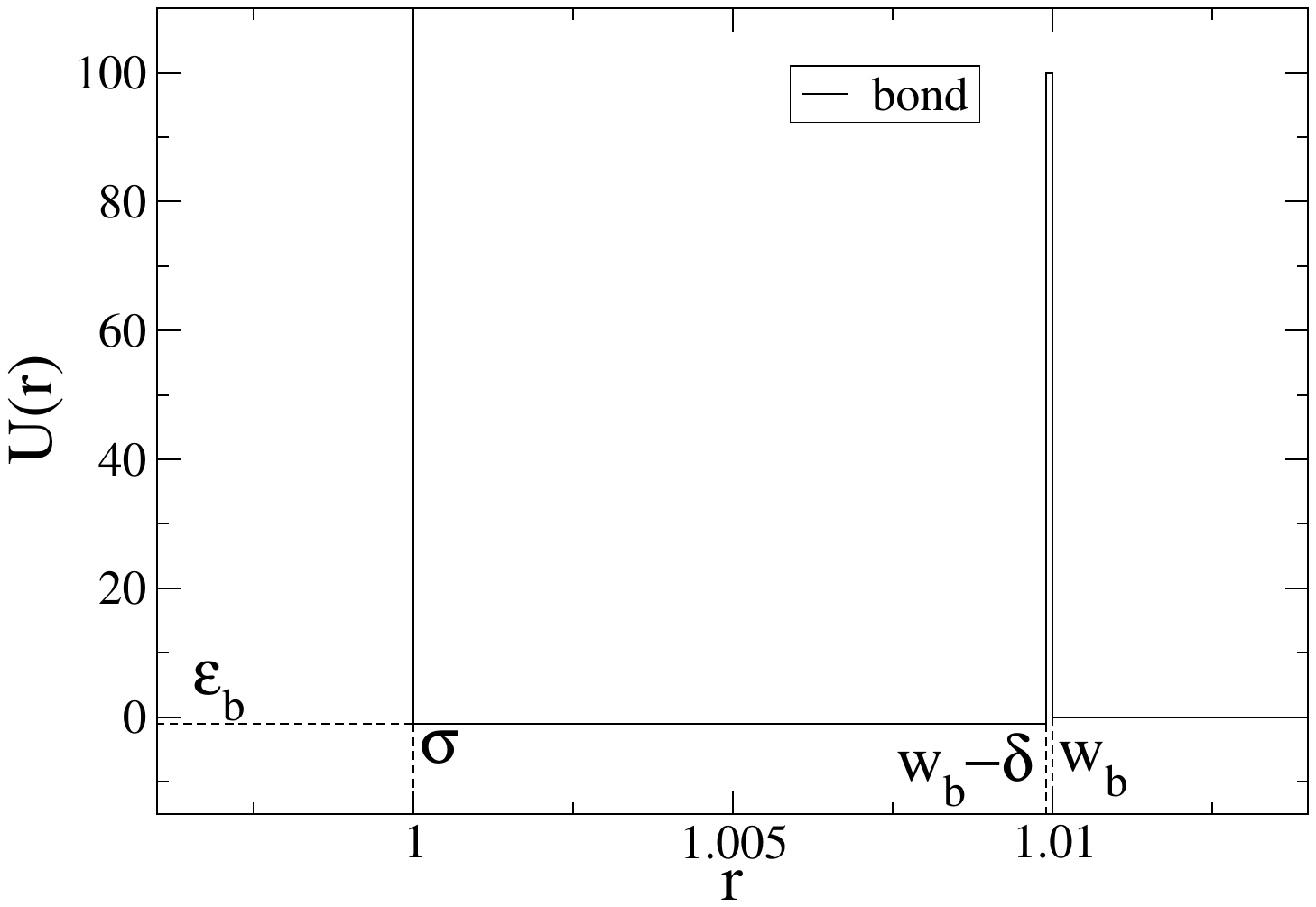}
    \caption{The potential of the bond between atoms of type 1 and 1 and atoms of 1 and 2. preventing atoms of type 1 to separate and become monomers. The height of the potential barrier can be taken as high as 1000 and the width $\delta$ can be taken as small as 0.0001.} 
    \label{fig:7}
\end{figure}

The maximum valence model for $z=2$ has been used to mimic the LLPT associated with polymerization~\cite{Shumovsky2022} as was experimentally observed in sulfur at high pressures~\cite{Henry_Sulfur_2020}. One of many unrealistic features of that model is that real sulfur remains molecular S$_2$ near its
liquid--gas critical temperature, while in the limited valence model the dimers at such a temperature dissociate into monomers. Here, we modify
the maximum valence model in its Hamiltonian formulation assigning the atoms with zero neighbors in their bond range an infinite potential
energy. Within the DMD package, this can be achieved by modifying a bond
potential between an atom with one bond and its neighbor which may have either one or two bonds a quasi-infinite very narrow square potential of barrier as shown in figure~\ref{fig:7}. In this new model, the only possible types of atoms  are atoms with one bond forming dimers or being at the end of a polymer chain and atoms with two bonds forming a chain. We tested that near the LGCP, the gaseous phase has only 5.5\% atoms with two bonds, which are the middle atoms of the trimers and no atoms with zero bonds.

\begin{figure}[h]
    \centering
    \includegraphics[width=0.49\linewidth]{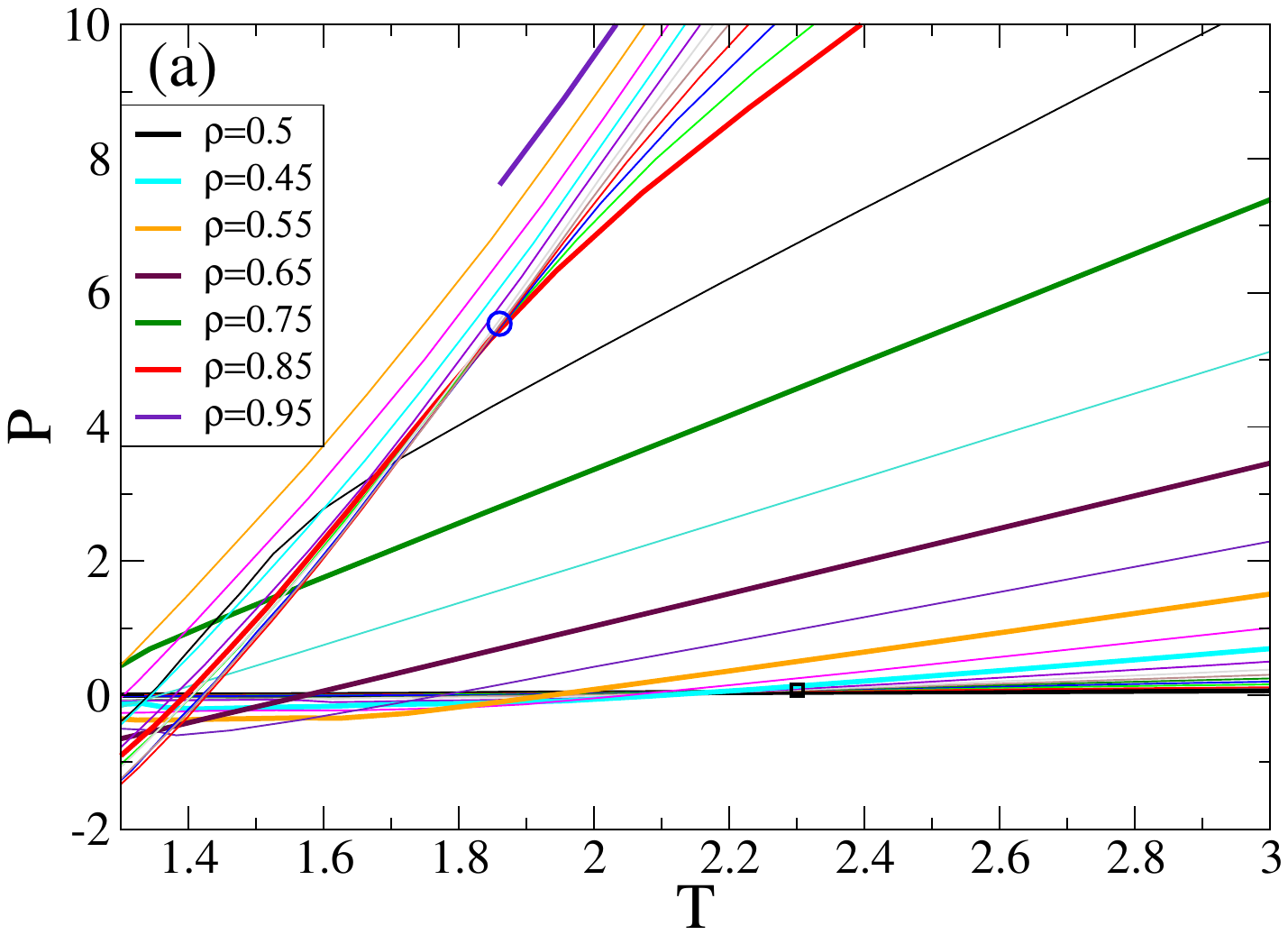}
    \includegraphics[width=0.49\linewidth]{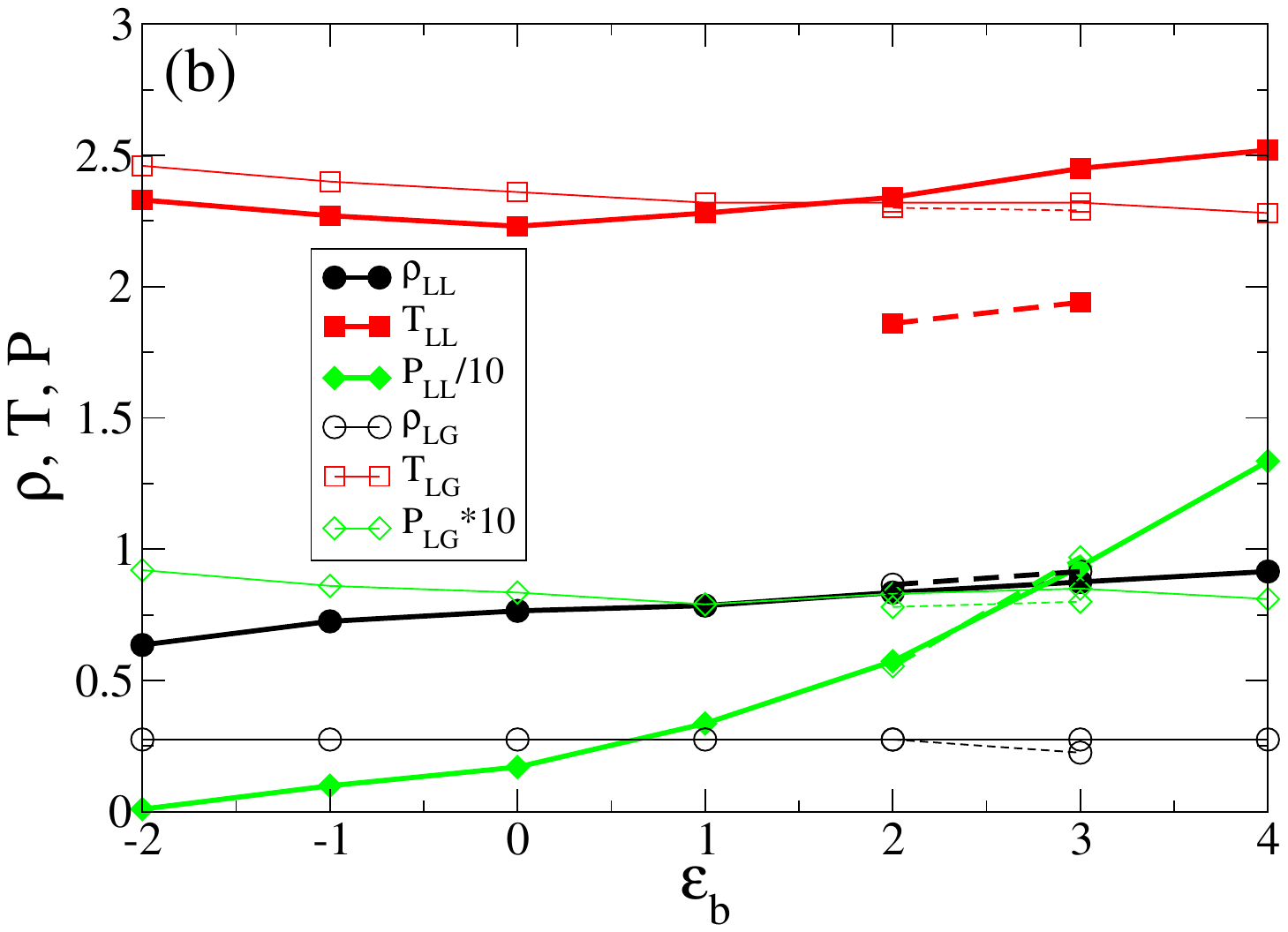}
    \caption{(Colour online) (a) The $P-T$ phase diagram for model of the polymerization of dimers ($z=2$, $w=1.7$, $w_b=0.01$, $w_z=1.3$, $\epsilon_b=2$ and $\epsilon_z=-1$). The lines are isochores with step $\Delta\rho=0.05$ from $\rho=0.05$ to
    $\rho=0.85$ and with step $\Delta\rho=0.01$ from $\rho=0.85$ to $\rho=0.95$. The LLCP is indicated by a circle while the LGCP
    is indicated by a square. The results are obtained by slow cooling with equillibration
    time for each point $1000$ time units. (b) The dependence of the critical point location
    on the repulsion/attraction energy of the bond $\epsilon_b$ for $z=2$, $w=1.7$, $w_b=0.01$, $w_z=1.3$, and $\epsilon_z=-1.3$. The values for LLCP (filled symbols) are connected with bold lines, while the values for LGCP
    (empty symbols) are connected with thin
    lines. One can see that changing bond from attractive to repulsive significantly increases $P_{LL}$, making the ratio $P_{LL}/P_{LG}\approx 100$ as in real sulfur for $\epsilon_b\approx 3$. Two additional points are given for $\epsilon_z=-1$, which yields a better agreement with the experimental data because reducing $\epsilon_z$ significantly reduces $T_{LL}$ making it below the $T_{LG}$ as in real sulfur, without affecting other critical values. These two
    points for $\epsilon_b=2$ and 3 are connected with dashed lines.}
    \label{fig:8}
\end{figure}

Qualitatively, the new Hamiltonian model with $z=2$ and forbidden monomers
have phase diagrams similar to the cases studied in reference~\cite{Shumovsky2022}. Here, as an example, we present a $P-T$ phase diagram for $w=1.7$, $w_b=0.01$, $w_z=1.3$, $\epsilon_b=2$ and $\epsilon_z=-1$ (figure~\ref{fig:8}). Note that the atoms belonging to polymer chains attract each other due to the additional attraction for atoms of type 2, but the cost of forming a bond is positive. Moreover, the bonds are very narrow. Technically, it is not the bonds which hold the polymer chain together but the additional attractive potential between the atoms with two bonds. These bonds with positive energy are not necessarily very unstable. They can be metastable with as long life time as we wish if we add to the bond potential a large enough potential barrier as in figure~\ref{fig:7}. In the limit of $\delta \to 0$, this potential will not change the Hamiltonian, and thus will not affect the phase diagram, but will dramatically slow down the kinetics.  Whether this type of bond potential can be justified by quantum chemistry remains an open question. 

Nevertheless, the model qualitatively describes the LLPT in sulfur [figure~\ref{fig:8}(a)]. It shows two phase transitions and their spinodals indicated by the loci of isochore
crossing which join at critical points: the liquid--liquid one 
indicated by a circle at $\rho_{LL}=0.865\pm 0.005$, 
$T_{LL}=1.86\pm 0.01$, and $P_{LL}=5.54\pm 0.05$ and a liquid--gas 
one indicated by a square at $\rho_{LG}=0.275\pm 0.025$, 
$T_{LG}=2.30\pm 0.05$, $P_{LG}=0.078$. Note that in real sulfur 
$T_{LG}=1314$~K, $P_{LG}=20.7$~MPa, and $\rho_{LG}= 563$~kg/m$^3$ 
\cite{CRC}, while the LLCP is located at $T_{LL}=1035$~K, $P_{LL}= 2.15$~GPa, and $\rho_{LL}\approx 2000$ 
kg/m$^3$~\cite{Henry_Sulfur_2020}, making the ratios of critical
parameters for real sulfur equal to $\rho_{LL}/\rho_{LG}=3.55$, 
$T_{LL}/T_{LG}=0.79$, $P_{LL}/P_{LG}=104$, while for the model 
with the given set of parameters they are, respectively, 3.2,  0.81
and 7.1. As one can see from figure~\ref{fig:8}(b) a further increase
of $\epsilon_b$, can easily increase both $P_{LL}/P_{LG}$ and  $\rho_{LL}/\rho_{LG}$.
Note also that the liquid--liquid coexistence line must belong to a narrow region between the spinodals and thus must have a very large positive slope. Moreover, it crosses the
liquid--gas coexistence line  at a very small temperature
and at a small pressure almost equal to zero,
where crystallization
can be anticipated, making the phase diagram very realistic, because according to reference~\cite{Henry_Sulfur_2020}, the liquid--liquid coexistence line in sulfur ends at a triple point with the crystalline phase.  

In order to improve the agreement with the experimental data on sulfur, we explore how the location of the critical
points is affected by the bond strength $\epsilon_b\equiv \Delta H_0$ (enthalpy of bond formation), figure~\ref{fig:8}.
We see that for negative bond formation enthalpy (exothermic reaction) leads to a very low
critical pressure of the LLCP making it even negative if $\epsilon_b<-2$. As $\epsilon_b$ grows, $P_{LL}$ increases and for $\epsilon_b\approx 3$ it becomes 100 times larger than
$P_{LG}$ as in real sulfur. When $\epsilon_b$ increases, the critical density, $\rho_{LL}$, also  increases from $0.63$ at $\epsilon_b=-2$ to $0.915$ at $\epsilon_b=4$. The critical temperature $T_{LL}$ slowly increases with $\epsilon_b$, becoming greater than $T_{LG}$ for $\epsilon_b>3$, which is not correct for real sulfur.
The way to overcome this problem is to increase $\epsilon_z$ from $-1.3$ to $-1.0$. This significantly reduces $T_{LL}$ without any significant effect on other critical parameters. The parameters of the LGCP do not
depend on $\epsilon_b$ and $\epsilon_z$ in any significant way.

\section{Conclusion}
We develop a Hamiltonian variant of the limited valence model in which each atom is represented by a single particle and implement it via the event driven molecular dynamics. The new model has a well-defined Hamiltonian, which allows it to be treated within
the formalism of statistical mechanics. For example, it has a well-defined entropy and can be studied by the Monte Carlo method. We show that this model provides a simple way to reproduce LLPT in various substances. The six independent parameters of the model $z$,
$w$, $w_b$, $w_z$, $\epsilon_b$, and $\epsilon_z$ strongly change the location of the LLPT and its interplay with the LGPT. We also study the relation of this new model with the previously studied core-shell non-Hamiltonian limited valence model of references~\cite{Shumovsky2022,Shumovsky2023} and show that qualitatively all the phenomena observed there are reproduced by the new model, which is much faster computationally.

For $z=1$, we show that the model can reproduce molecular-metallic transition in liquid hydrogen. For $z=2$, it can reproduce the LLPT associated with polymerization of dimers in sulfur. And for $z=4$, it can reproduce the phase diagram
of water with a LLPT at low temperature and high pressure. The limited valence model is thus a powerful, versatile and simple tool to study LLPTs. We hope that in future it can be applied to the LLPT in phosphorus for $z=3$ and in many other compounds.

We find that for $z=1$ and narrow repulsive shoulder $w_z=1.15$, it is possible to simultaneously observe the LLCP and the solid--solid isostructural critical point at slightly different densities and temperatures.

For the case of $z=4$, we find that for the large values of $w_z\approx 1.34$ which can be relatated to the position of the minimum of the oxygen-oxygen radial distribution function in water, the phase diagram of the model resembles the phase diagram of water obtained by all-atom simulations.  

For the case of sulfur ($z=2$) we modify the limited valence model to also include the minimum valence $v_{\rm min}=1$, forbidding the formation of monomers.

\section*{Acknowledgement}
I am grateful to the administration of the Boston University Department of Physics for providing me the research computing support. I also acknowledge the support by NSF Award No. 1856496. I am grateful to Marina Eskin for proofreading the manuscript and to Mikhail A. Anisimov, Thomas J. Longo, Nikolay Shumovskyi  and Francesco Sciortino for useful discussions and input at the initial stages of this work.

\newpage
\ukrainianpart

\title{Гамільтоніан-обмежені валентні моделі для поліморфізму рідин}

\author{С. В. Булдирев\refaddr{label1, label2}}
\addresses{
	\addr{label1} Фізичний факультет, Університет Єшиви, Нью-Йорк, NY 10033, США
	\addr{label2} Фізичний факультет, Бостонський університет, MA 02215, США  
}

\makeukrtitle

\begin{abstract}
	\tolerance=3000%
	Фазові переходи ``рідина--рідина'' знайдені експериментально або з використанням комп'ютерного моделювання у багатьох сполуках, таких як вода, водень, сірка, фосфор, вуглець, кремнезем та кремній. Обмежена валентна модель, застосована через алгоритм керованої подіями молекулярної динаміки, забезпечує простий загальний механізм для фазових переходів ``рідина--рідина'' в усіх цих різноманітних випадках. У статті представлено один варіант обмеженої валентної моделі з добре визначеним гамільтоніаном, а саме: унікальний алгоритм, з допомогою якого потенціальну енергію системи частинок можна обчислити виключно з їх координат, що є еквівалентним до потенціалу багатьох частинок. Представлено декілька прикладів моделі, які можна застосовувати для відтворення фазового переходу ``рідина--рідина'' в системі з максимальними валентностями $z=1$ (водень), $z=2$ (сірка) та $z=4$ (вода), де $z$ --- найбільше число зв'язків, які може мати атом. Для $z=1$ встановлено низку параметрів, при яких дана система має кри\-тич\-ні точки ``рідина--рідина'' та ізоструктурні критичні точки ``тверде тіло--тверде тіло''. Для $z=4$ знайдено параметри, при яких фазова діаграма подібна на фазову діаграму води з широкою областю від'ємних коефіцієнтів теплового розширення (густинних аномалій), яка поширюється на метастабільну область від'ємного тиску. Модель обмеженої валентності може бути модифікована для того, щоб заборонити не тільки занадто великі валентності, але й досить малі. У випадку сірки забороняється утворення мономерів, що обмежує валентність $v$ атома в межах інтервалу $1=v_{\rm min}\leqslant v\leqslant v_{\rm max}\equiv z=2$.
	\keywords фазові переходи рідина--рідина, рідина--газ, перетворення тверде тіло--тверде тіло, фазові діаграми
	
\end{abstract}

\lastpage
\end{document}